\documentclass[aps,
prl,
nofootinbib,
floatfix,
twocolumn,
superscriptaddress,
showpacs
]{revtex4-1}

\usepackage{graphicx}% Include figure files
\usepackage{color} % use colored fonts
\usepackage{dcolumn}% Align table columns on decimal point
\usepackage{bm}% bold math
\usepackage{epsfig}
\usepackage[subfigure]{graphfig}
\usepackage{latexsym}
\usepackage{soul}
\usepackage{amsmath}
\usepackage{amsfonts}
\usepackage{amsxtra}

\newcommand{\lwrsim}{\raise0.3ex\hbox{$<$\kern-0.75em\raise-1.1ex\hbox{$\sim$}}}
\def\krto{ {\,\,\lower .8ex\hbox {$\longrightarrow \atop k \rightarrow 0$}\,\,}}

\def\bea{\begin{eqnarray} }
\def\beq{\begin{eqnarray} }

\def\eea{\end{eqnarray}}
\def\eeq{\end{eqnarray}}

\begin{document} %\date{\today}
\null\hfill\begin{tabular}[t]{l@{}}
  {JLAB-THY-20-3161}
\end{tabular}
\preprint{JLAB-THY-20-3161}
%\date{}
%%%%%%%%%%%%%%%%%%%%%%%%%%%%%%%%%%%%%%%%%%%%%%
%% title page 
%%%%%%%%%%%%%%%%%%%%%%%%%%%%%%%%%%%%%%%%%%%%%%

\title{Parton Distribution Functions from Ioffe Time Pseudodistributions from Lattice Calculations: Approaching the Physical Point}

\newcommand*{\ORNL}{Oak Ridge Leadership Facility, Oak Ridge National Laboratory,
One Bethel Valley Road, Oak Ridge, TN 37831, USA}\affiliation{\ORNL}
\newcommand*{\JLAB}{Thomas Jefferson National Accelerator Facility, Newport News, VA 23606, USA}\affiliation{\JLAB}
\newcommand*{\CU}{Physics Department, Columbia University, New York City, New York 10027, USA}\affiliation{\CU}
\newcommand*{\WM}{Physics Department, College of William and Mary, Williamsburg, Virginia 23187, USA}\affiliation{\WM}
\newcommand*{\ODU}{Physics Department, Old Dominion University, Norfolk, VA 23529, USA}\affiliation{\ODU}
\newcommand*{\Marseille}{Aix Marseille Univ, Universit\'e de Toulon, CNRS, CPT, Marseille, France}\affiliation{\Marseille}

\author{B\'alint Jo\'o}\affiliation{\ORNL}\affiliation{\JLAB}
\author{Joseph Karpie}\affiliation{\CU}
\author{Kostas Orginos}\affiliation{\JLAB}\affiliation{\WM} 
 \author{\mbox{Anatoly V. Radyushkin}}\affiliation{\JLAB}\affiliation{\ODU}
\author{David G. Richards}\affiliation{\JLAB}
\author{Savvas Zafeiropoulos}\affiliation{\Marseille}
\collaboration{On behalf of the \textit{HadStruc Collaboration}}

\begin{abstract}
We present results for the unpolarized parton distribution function of the nucleon computed in lattice QCD at the physical pion mass. This is the first study of its kind employing the method of Ioffe time pseudo-distributions. Beyond the reconstruction of the Bjorken-$x$ dependence we also extract the lowest moments of the distribution function using the small Ioffe time expansion of the Ioffe time pseudo-distribution. We compare our findings with the pertinent phenomenological determinations.

\end{abstract}


\maketitle

%\vfill
%\newpage

%%%%%%%%%%%%%%%%%%%%%%%%%%%%%%%%%%%%%%%%%%%%%%%%%%%%%%%%%
%% end of title page
%%%%%%%%%%%%%%%%%%%%%%%%%%%%%%%%%%%%%%%%%%%%%%%%%%%%%%%%%

%%%%%%%%%%%%%%%%%%%%%%%%%%%%%%%%%%%%%%%%%%%%%%%%%%%%%%%%%
%% body of the paper
%%%%%%%%%%%%%%%%%%%%%%%%%%%%%%%%%%%%%%%%%%%%%%%%%%%%%%%%%

%\section{Introduction}
%\alinea

\noindent\emph{Introduction}.\,---\,
The determination and understanding of the internal quark and gluon structure of the proton is a crucial aspect of the precision phenomenology program of the current and future hadron collider experiments, especially the Large Hadron Collider (LHC) and the upcoming Electron-Ion Collider (EIC). The framework of collinear factorization quantifies the hadronic structure in terms of Parton Distribution Functions (PDFs) which encapsulate the pertinent information regarding the momentum distributions of quarks and gluons within the nucleon. Till very recently, the intrinsic non-perturbative nature of the PDFs was prohibiting an ab-initio computation and the conventional approach is to employ a variety of experimental data together with advanced fitting methodologies in order to extract the PDFs via global fits. The studies of PDFs are of paramount importance precisely due to the fact that their uncertainties play a crucial role in many LHC applications. They affect the measurement of precision SM parameters, such as the $W$ mass, the strong coupling constant and the determination of the couplings of the Higgs boson where discrepancies from the stringently fixed SM predictions would serve as indisputable evidence of BSM physics~\cite{Gao:2017yyd}.

The possibility to determine the PDFs with first principle lattice calculations is the object of a long endeavor which recently lead to a culmination of results. The primary difficulty impeding a first principle implementation is associated with the fact that the matrix elements defining  the PDFs involve  light-cone separated fields. In his seminal article that stimulated the recent efforts,  X. Ji~\cite{Ji:2013dva} proposed to compute matrix elements of
fields separated by a purely space-like distance $z=z_3$ 
  that define the so-called quasi-PDF,
 the distribution in the longitudinal momentum $p_3$.
In  the large 
$p_3$ limit, they
 can be factorized into the light-cone PDF, $f(x,\mu^2)$. Subsequently, many articles studying quasi-PDFs, as well as the pion
quasi-distribution amplitude (DA) appeared in the literature~\cite{Lin:2014zya,Chen:2016utp,Alexandrou:2015rja,Alexandrou:2016jqi, Monahan:2016bvm, Zhang:2017bzy, Alexandrou:2017huk, Green:2017xeu, Stewart:2017tvs,Monahan:2017hpu, Broniowski:2017gfp,Alexandrou:2018pbm,Chen:2018xof,Alexandrou:2018eet,Lin:2018qky,Fan:2018dxu,Liu:2018hxv,Alexandrou:2019lfo,Izubuchi:2019lyk,Green:2020xco,Chai:2020nxw,Lin:2020ssv}.

 Alternative approaches 
 based on the analysis of equal-time current correlators 
 ~\cite{Detmold:2005gg,Braun:2007wv,Chambers:2017dov, Liang:2019frk} also  aim to study the PDFs or DAs in lattice QCD. 
``Good Lattice Cross-Sections'' (LCS),  
  as described  
in~\cite{Ma:2017pxb}, represent a general framework,  where one computes matrix elements that can be factorized into PDFs at short distances. Works of~\cite{Bali:2017gfr,Bali:2018spj,Sufian:2019bol,Bali:2019ecy, Sufian:2020vzb} fall into these categories. 
For comprehensive reviews on the topic, we refer the reader to ~\cite{Lin:2017snn, Cichy:2018mum, Monahan:2018euv, Qiu:2019kyy}.

\smallskip
\noindent\emph{Ioffe time pseudo-distributions}.\,---\,
Another position-space formulation was proposed in~\cite{Radyushkin:2017cyf}. In this 
approach,  the basic object is 
the Ioffe time pseudo-distribution function (pseudo-ITD)  $\mathcal{M}(\nu,z^2)$.
The Lorentz invariant $\nu=p \cdot z$
 is known as the Ioffe time~\cite{Ioffe:1969kf,Braun:1994jq}. The pseudo-ITD is the 
invariant 
amplitude for 
a matrix element with space-like separated quark fields. 

In renormalizable theories, the pseudo-ITD 
exhibits a logarithmic 
singularity 
at small values of $z^2$.  These short-distance 
singularities  can be factorized into the PDF and a perturbatively calculable coefficient function.  The pseudo-ITD can also be considered as a LCS.  A series of works implemented this formalism and studied its efficiency~\cite{Orginos:2017kos,Karpie:2017bzm,Karpie:2018zaz,Karpie:2019eiq,Joo:2019jct,Joo:2019bzr}. For the sake of completeness, the main points of our formalism are summarized below, but we refer the reader to~\cite{Joo:2019jct,Radyushkin:2019mye} for a detailed discussion.

The non-local matrix element,
\begin{equation} \label{eq:a_matrix_element} 
M^\alpha (p,z) = \langle p | \bar\psi(z) \gamma^\alpha U(z;0) \psi(0) | p \rangle \,,\end{equation}
with  
$U$ being  a straight Wilson line,  
 $p = (p^+, \frac{m^2}{2p^+},0_T)$,  $z = (0,z_- , 0_T)$  and $\gamma^a= \gamma^+$ in light-cone coordinates,  defines the ${\overline{\rm MS}}$ ITD
(introduced in \cite{Braun:1994jq}), 
 given a regularization
 is made 
  for the $z^2=0$ 
  singularity.  
 For $z^2 \neq 0$, this 
matrix element has the following Lorentz decomposition
\begin{equation} \label{eq:pseudo_lor_decomp}
M^\alpha(z,p) = 2 p^\alpha \mathcal{M}(\nu,z^2) + 2 z^\alpha\mathcal{N}(\nu,z^2) \, .
\end{equation}
 %%%%%  where the Lorentz invariant $\nu=p\cdot z$ is called the Ioffe time.  
The pseudo-ITD  
$\mathcal{M}(\nu,z^2)$ contains 
the leading twist contribution, 
 while  
$\mathcal{N}$
is a higher-twist  term.   
In the  kinematics  \mbox{$p=(E,0,0,p_3)$, $z=(0,0,0,z_3)$,}
the choice
$\alpha=0$ isolates $\mathcal{M}$. Nonetheless, it still contains higher twist contaminations $O(z^2 \Lambda^2_{\rm QCD})$. In the limit of small $z^2$, where higher twist terms are suppressed, 
 $\mathcal{M}$ is factorizable into the ITD  (or equivalently,  the PDF) and a perturbative coefficient function, provided that 
 one  removes 
Wilson line-related 
UV divergences that appear  at finite $z^2$. 
These UV divergences are eliminated if one 
considers  the reduced pseudo-ITD 
~\cite{Radyushkin:2017cyf,Orginos:2017kos}   
 given by the ratio 
 \beq\label{eq:redu}\mathfrak{M}(\nu,z^2) = \frac{\mathcal{M}(\nu,z^2)}{\mathcal{M}(0,z^2)} \,.\eeq
 It contains the same singularities 
 in the $ z^2=0$ limit as $ \mathcal{M}$,  and can be related 
 to the $\overline{\rm MS}$ light-cone ITD, $Q(\nu,\mu^2)$, by 
  the  NLO matching relation ~\cite{Radyushkin:2018cvn,Zhang:2018ggy,Izubuchi:2018srq}

\begin{align} 
\label{evolution}
\mathfrak{M}(\nu,z^2) =& Q(\nu,\mu^2) 
-  
\frac{\alpha_s C_F}{2\pi} \int_0^1 du \, Q(u\nu,\mu^2)  
\times \nonumber \\ &  
\biggl [ \ln\left  (z^2\mu^2 \frac{e^{2\gamma_E +1}}4 \right )
B(u) +L(u) \biggr  ] \,,
\end{align} 
where $B(u) = \left[ \frac{1+u^2}{1-u}\right]_+$ is the Altarelli-Parisi kernel~\cite{Altarelli:1977zs},  and
 \beq
 \label{eq:lu}L(u) = 
  \left[ 4 \frac{\ln(1-u)}{1-u} - 2(1-u) \right]_+\,.
  \eeq

%------ensembles
\begin{table*}[t] 
\centering
\begin{tabular}{ l | c c | c c | c c | c c }
ID & ~$a$(fm)~ & ~$M_\pi$(MeV) & ~$\beta~$ & ~$c_{\rm SW}$~ & ~$am_l$~ & ~$am_s$~ & $L^3 \times T$ & $N_{\rm cfg}$\\\hline\hline
$a094m360$ & 0.094(1) & 358(3) & 6.3 & 1.20536588 & -0.2350 & -0.2050 & $32^3 \times 64$ & $417$\\\hline
$a094m280$ & 0.094(1) & 278(3) & 6.3 & 1.20536588 & -0.2390 & -0.2050 & $32^3 \times 64$ & $500$ \\\hline
$a091m170$ & 0.091(1) & 172(6) & 6.3 & 1.20536588 & -0.2416 & -0.2050 & $64^3 \times 128$ & $175$\\\hline \hline\end{tabular}
\caption{\label{tab:lat}\footnotesize Parameters for the lattices generated by the JLab/W\&M collaboration using 2+1 flavors of stout-smeared clover Wilson fermions and a tree-level tadpole-improved Symanzik gauge action. More details about these ensembles can be found in~\cite{Yoon:2016jzj}. }
\end{table*}
%---------------------------------
\smallskip
\noindent\emph{Extracting the matrix element.}\,---\,
The numerical  computation  of our matrix elements relies on  Gaussian smearing~\cite{Allton:1993wc} and momentum-smearing~\cite{Bali:2016lva}  for constructing the nucleon interpolating field, as well as the summation method for better control of the excited state contamination.  The latter is intimately related to the Feynman-Hellmann (FH) theorem~\cite{Bouchard:2016heu} and has been widely used in Lattice calculations of PDFs~\cite{Orginos:2017kos,Karpie:2017bzm,Fan:2018dxu,Alexandrou:2019lfo,Izubuchi:2019lyk,Joo:2019jct,Joo:2019bzr}.

The matrix element is determined from a ratio of correlation functions
\bea
R(t) = \frac{\displaystyle \sum_\tau C_3(t,\tau)}{C_2(t)} \, ,
\eea
where $C_{2,3}$ are standard two and three point correlation functions, $t$ is the Euclidean separation between the source and sink interpolating fields, and the operator insertion time $\tau$ is summed over the entire temporal range. The effective matrix element $M^{\rm eff}$ is then constructed as
\bea
M^{\rm eff} (t) = R(t+1) - R(t)\,.
\eea
The leading excited-state effects can be parameterized by
\bea
\label{eq:fitform}M^{\rm eff}(t) = M ( 1 + A e^{-\Delta t} + B t e^{-\Delta t}) \,.
\eea
with $\Delta$ being the energy gap between the ground state and the lowest excited state.

The summation method has a clear advantage over the typical ratio method. The excited state contamination scales as ${\rm exp}(-\Delta t)$ instead of ${\rm exp}(-\Delta t/2)$, which allows for smaller $t$ to be used to control excited state effects. Since correlation functions' errors grow exponentially, the summation method requires significantly fewer measurements to obtain a desirable statistical precision for data with controlled excited states. This feature is important for calculations  at large momenta, where energy gaps can be small and the error decays much faster than for low momenta.  

\begin{figure}[!htp]
\centering
\subfigure[]{\includegraphics[width=0.48\textwidth]{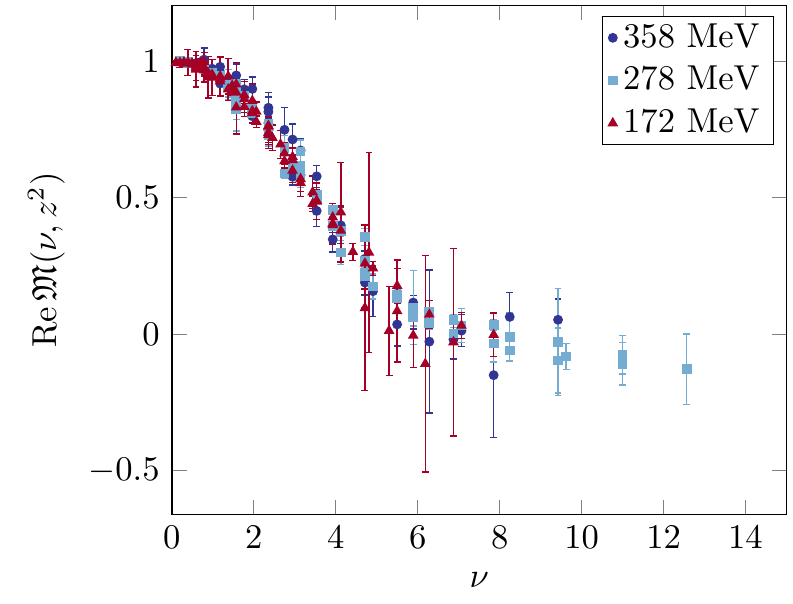}\label{runmat}}
\subfigure[]{\includegraphics[width=0.48\textwidth]{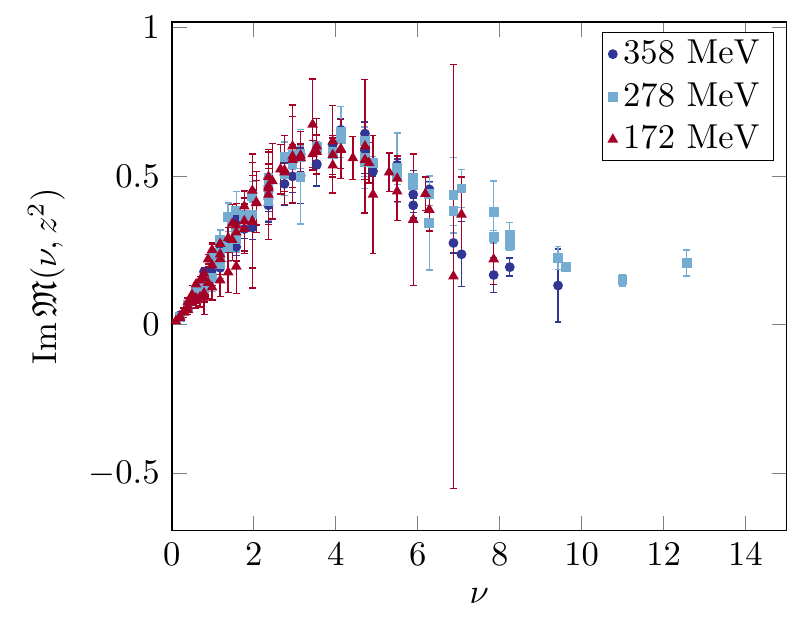}\label{inmat}}
\caption{\label{fig:pitd0} The reduced pseudo-ITD calculated on ensembles with 358 MeV, 278 MeV, and 172 MeV pion masses. The upper and lower plots are the real and imaginary component respectively. There appears to be very small mass effects within this range of $\nu$ and $z^2$. }
\end{figure}

\smallskip
\noindent\emph{Lattice QCD calculation.}\,---\,
In this study, three ensembles of  configurations with decreasing value of the pion mass have been employed.  In Tab.~\ref{tab:lat}, we list all the parameters of our analysis. 
The pion masses of this study are 172 MeV, 278 MeV, and 358 MeV.  These ensembles allow for a controlled extrapolation to the precise physical pion mass which constitutes an important limit to be taken in order to safely compare with the PDF determinations of global fits but also for the first time we can study the pion mass effects on the ITD.
  As was done in~\cite{Joo:2019jct}, correlation functions with several different smearings were simultaneously fit to determine the matrix element from Eq.~\eqref{eq:fitform}.
 The matrix elements extracted from fitting correlation functions to Eq.~\eqref{eq:fitform} are shown in Fig.~\ref{fig:pitd0}.
 
 %---------------------------------
%\section{moments} 

\begin{figure}[!htp]
\includegraphics[width=0.48\textwidth]{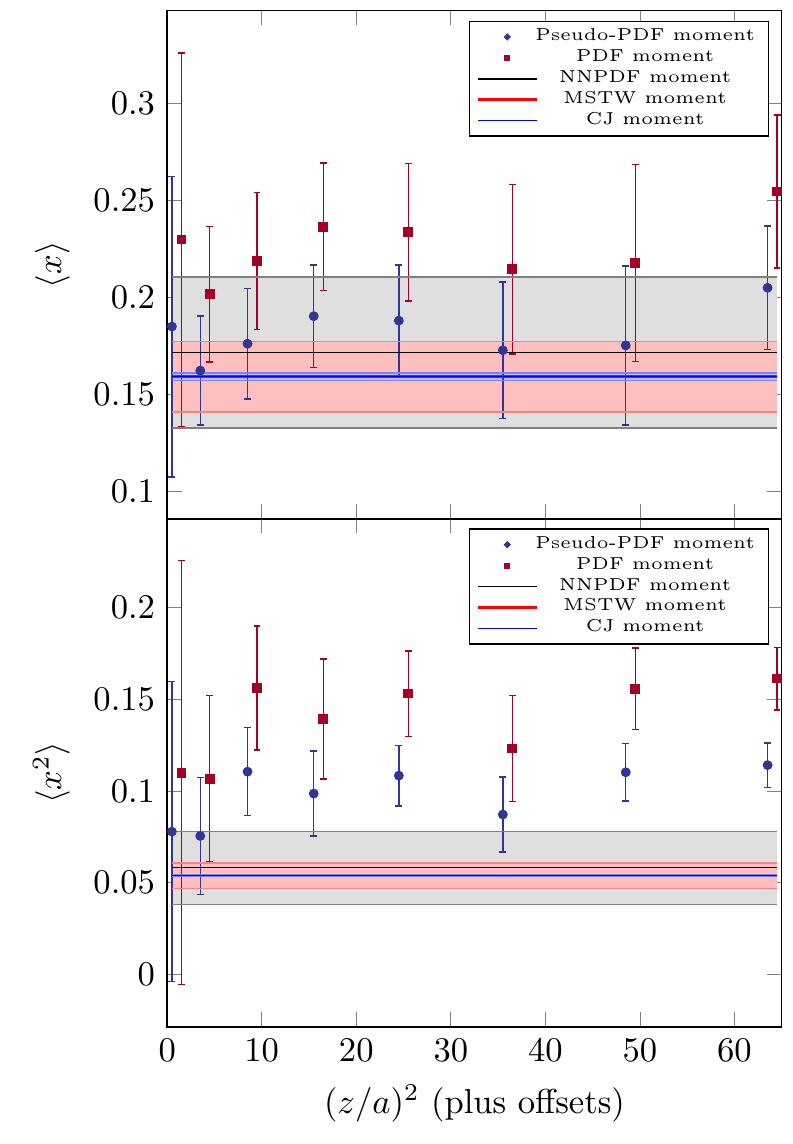}
\caption{\label{moms}
The first two moments of the pseudo and the $\overline{\rm MS}$ lightcone PDF computed from the ensemble $a091m170$, compared to phenomenologically determined PDF moments from the NLO global fit CJ15nlo~\cite{CJ}, the NNLO global fits MSTW2008nnlo68cl\_nf4~\cite{Martin:2009iq} and NNPDF31\_nnlo\_pch\_as\_0118\_mc\_164~\cite{Ball:2017nwa} all evolved to 2 GeV. }
\end{figure}

\smallskip
\noindent\emph{Moments of the PDF.}\,---\, Following our suggestion in~\cite{Karpie:2018zaz}, we can use the reduced pseudo-ITD to compute the moments of the PDF. Valuable information
for the PDF can be extracted from the data without dealing with the pitfalls of the inverse problem.
The moments of the $\overline{\mbox{MS}}$ PDF, $a_n(\mu^2)$,  are related multiplicatively to those of the Fourier transform of the reduced pseudo-ITD,
\begin{equation}
\label{eq:mom_matching}
 b_n(z^2) = C_n(\mu^2z^2) a_n(\mu^2) + \mathcal{O}(z^2 \Lambda^2_{\rm QCD})
 \end{equation}
 where $C_n$ are the Mellin moments of the matching kernel $C(u,\mu^2z^2)$ with respect to $u$.
 To NLO accuracy, 
 \begin{equation}
 \label{eq:pmom_match}
 C_n(z^2\mu^2) = 1 - \frac{\alpha_s}{2\pi} C_F \left[\gamma_n \ln\left(z^2\mu^2\frac{e^{2\gamma_E +1}}{4}\right) + l_n\right]\,, 
 \end{equation}
 where
 \begin{equation}
  \gamma_n = \int_0^1 du\, B(u) u^n= \frac{1}{(n+1) (n+2) } - \frac{1}{2} - 2 \sum_{k=2}^{n+1}\frac{1}{k}\,, 
  \end{equation} 
  are the moments of the Altarelli-Parisi kernel, and
\begin{align}    
     l_n =   \int_0^1 du\, L(u) u^n 
    = & 2\left[ \left(\sum_{k=1}^n \frac{1}{k}\right)^2 + \sum_{k=1}^n \frac{1}{k^2}  \right. \nonumber \\ & \left. 
+\frac12 - \frac{1}{(n+1)(n+2)} \right] .   
 \end{align}   
The even and odd moments can be determined from the coefficients of polynomials which are fit to the real and imaginary components respectively. The order of the polynomial is chosen to minimize the $\chi^2$/d.o.f. for each $z^2$ separately. As an example, the first and second moments calculated on the ensemble $a091m170$ are shown in Fig.~\ref{moms}. The $z^2$ dependence of the resulting PDF moments can be used to check for the size of higher twist effects, which do not seem significant. 
   
\begin{figure}[!htp]
\centering
\subfigure[]{\includegraphics[width=0.48\textwidth]{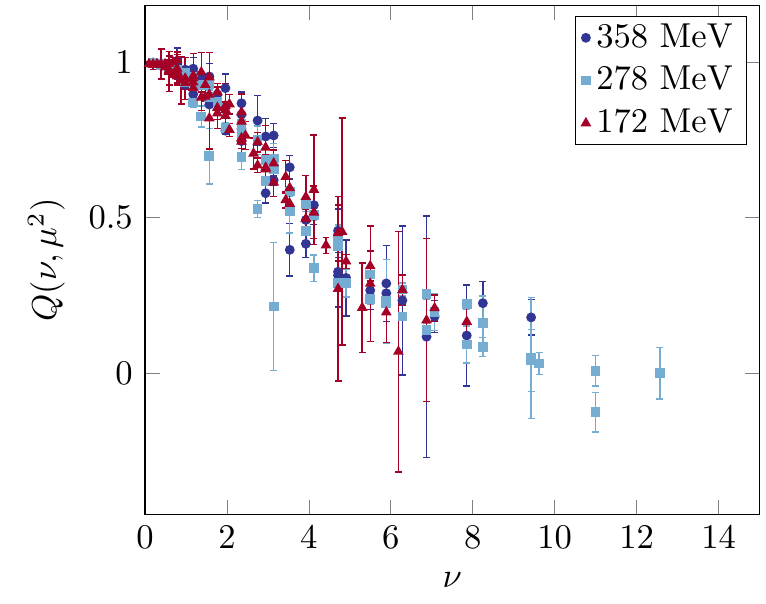}\label{fig:evo}}
\subfigure[]{\includegraphics[width=0.48\textwidth]{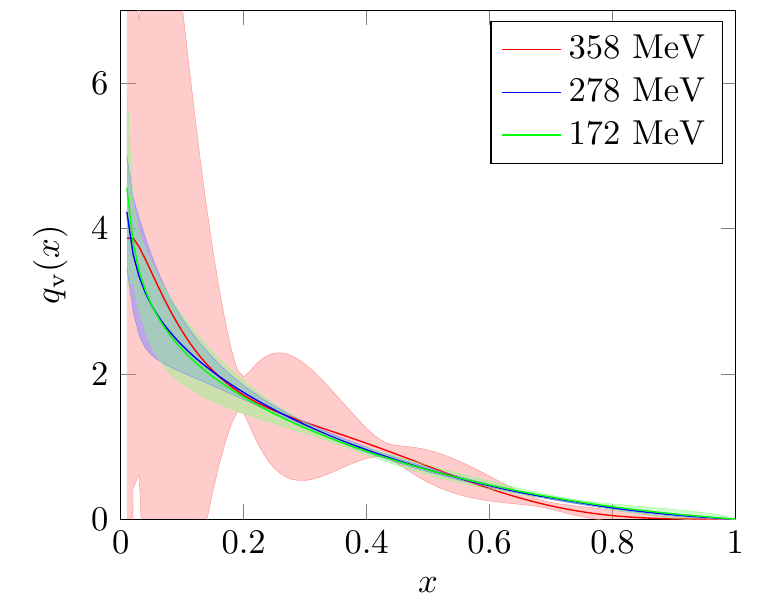}\label{fig:pdf}}
\caption{ (Upper) The $\overline{\rm \mbox MS}$ ITD matched to 2 GeV from the reduced pseudo-ITD results calculated at 358 MeV, 278 MeV, and 172 MeV. (Lower) The nucleon valence distribution obtained from fitting the ITD to the form in Eq.~\eqref{eq:pdf_fit_form} from each of those ensembles. }
\end{figure}

\smallskip
\noindent\emph{Matching to $\overline{ \mbox MS}.$}\,---\,
%---------------------------------
Similarly to Ref.~\cite{Joo:2019jct}, the reduced pseudo-ITD from each ensemble is matched to the lightcone $\overline{\rm MS}$ ITD at a given scale $\mu$ by inverting Eq.~\eqref{evolution}. As a result, we obtain a set of \mbox{$z^2$-independent}  curves for $Q(\nu,\mu^2)$ at $\mu= 2$ GeV, shown in Fig.~\ref{fig:evo}. 

As seen in the moments, the matching procedure has a small   \mbox{${\cal O}  (\alpha_s/\pi) \sim 0.1$} effect on the distribution. The contributions from the convolution of $B$ and $L$ with the reduced pseudo-ITD appear with opposite signs. The convolution with $L$ is slightly larger in magnitude, but by a factor which is approximately the same as the logarithmic coefficient of $B$. This feature may just be a coincidence at NLO, but it hints that higher order corrections may also be small. An NNLO or non-perturbative matching is required to check the effects of the perturbative truncation on the matching. 
 
\smallskip
\noindent\emph{Determination of the PDF.}\,---\,
The inversion of the Fourier transform defining the ITD, given a finite amount of data, constitutes an ill-posed problem which can only be resolved by including additional information. As was shown in~\cite{Karpie:2019eiq}, the direct inverse Fourier transform can lead to numerical artifacts, such as artificial oscillations in the resulting PDF. Many techniques have been proposed to accurately calculate PDFs from lattice data~\cite{Karpie:2019eiq,Izubuchi:2019lyk,Liang:2019frk,Cichy:2019ebf}. This issue also occurs 
in the determination of the PDF from experimental data. 

As was done in Ref.~\cite{Joo:2019jct}, the approach which is used here
(and is common amongst phenomenological determinations)  is to include information in the form of a model-dependent PDF parameterization. The parameterization used here is
  \beq
  \label{eq:pdf_fit_form}q_v(x) = \frac{1}N x^a (1-x)^b (1+ c\sqrt x + dx) \,, 
  \eeq  
where $N$ normalizes the PDF. The fits to this form, together with the bands representing the statistical errors on the fit, are shown in Fig~\ref{fig:pdf}.  In a future work, we will attempt to study the dependence on the choice of functional forms.
 
The results of these fits are largely consistent with each other. The heaviest pion mass PDF has notably larger statistical error than the others. This effect is due to a larger variance in the highly correlated $c$ and $d$ parameters. In the lighter two pion masses, the correlation between these parameters appears stronger, leading to a smaller statistical error in the resulting PDFs.

\begin{figure}[!htp]
\centering
\subfigure[]{\includegraphics[width=0.48\textwidth]{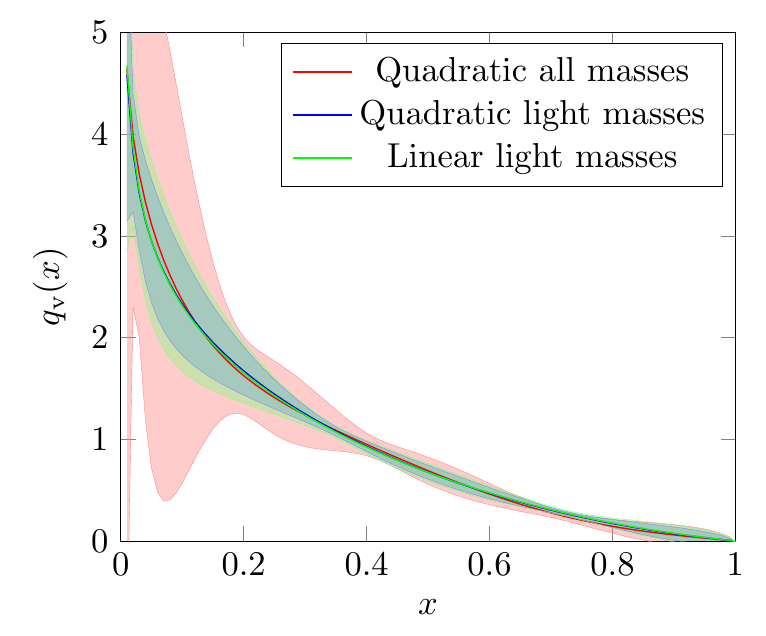}\label{fig:pdf_extrap}}
\subfigure[]{\includegraphics[width=0.48\textwidth]{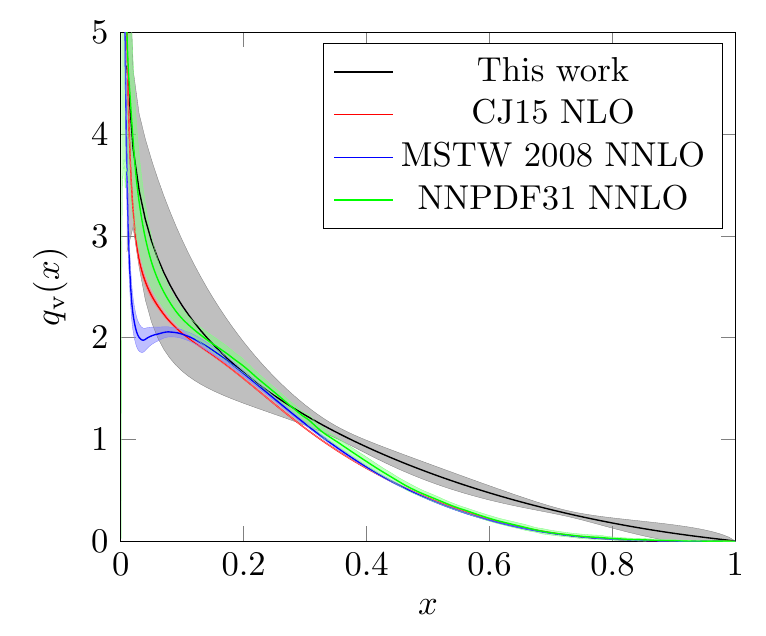}\label{fig:pdf_pheno}}
\caption{ (Upper) The extrapolations of the nucleon valence PDF to physical pion mass. (Lower) The nucleon valence distribution  compared to phenomenological determinations from the NLO global fit CJ15nlo~\cite{CJ} (green), and the NNLO global fits MSTW2008nnlo68cl\_nf4~\cite{Martin:2009iq} (red) and NNPDF31\_nnlo\_pch\_as\_0118\_mc\_164~\cite{Ball:2017nwa} (blue) at a reference scale of 2 GeV. }
\end{figure}

\smallskip
\noindent\emph{Extrapolation to the physical pion mass.}\,---\,
In order to determine the valence PDF for physical pion mass,  our results  must be extrapolated to 135 MeV. To do this, the central values of these curves are extrapolated and the errors are propagated. We have performed the extrapolation including and excluding the statistically noisy result from the heaviest pion ensemble. When using all three ensembles, we extrapolate the results using the form
\beq
q_v(x,\mu^2, m_\pi) = q_v(x,\mu^2, m_0) + a \Delta m_\pi + b\Delta m_\pi^2\,,
\eeq
where $\Delta m_\pi=m_\pi - m_0$ and $m_0$ is the physical pion mass. When using only the two lighter pion mass ensembles, we fix either $a$ or $b$ to be zero. Though, these extrapolations are not guaranteed to satisfy the normalization of the PDF, we have found them to be close within statistical precision. The extrapolated PDFs are shown in Fig~\ref{fig:pdf_extrap}. The linear extrapolation with the lightest two ensembles is compared to phenomenological determinations in Fig~\ref{fig:pdf_pheno}. In both figures, the error-bands represent only the statistical error. 

The  PDF obtained from this fit, for $x \gtrsim 0.2$ is  larger than the phenomenological fits. 
 This feature is consistent with the larger value of the second moment compared to the global fits in Fig.~\ref{moms}. Other remaining systematic errors could explain this discrepancy. In this study, no attempt was made to remove higher twist effects. Though the estimation of low moments, which relies on low $\nu$, show no significant sign of higher twist effects, they could still be present at larger $\nu$ where the ITD becomes more sensitive to higher moments. Also, this calculation was performed on ensembles with a fairly coarse lattice spacing and uses data with $ap \sim O(1)$. Discretization errors have been shown~\cite{Joo:2019jct} to be potentially significant. Future calculations at smaller lattice spacings are required to control these effects. There also exist potentially notable finite volume corrections which may need to be controlled.

\smallskip
\noindent\emph{ Conclusions}.\,---\,
We presented the first calculation of the nucleon PDF 
based on the method of Ioffe time pseudo-distributions performed 
at the physical pion mass.
This was an important 
step that had to be taken in order to have a more meaningful comparison with the pertinent phenomenological results. 
Also,  
by studying three 
ensembles with different pion masses,  
we were  able to 
investigate 
the dependence of the ITD on the pion mass.
  We saw that it is relatively mild compared to expectations stemming from 
  the studies  of 
  $\langle x \rangle $~\cite{Constantinou:2014tga} and 
 calculations of quasi-PDFs~\cite{Alexandrou:2018pbm}. 

Compared to similar studies,  our analysis capitalizes 
on three key factors. First, the ratio of matrix elements that 
yields a clean way to avoid all pitfalls and systematics of fixed gauge non-perturbative renormalization. Second, the short distance factorization, that  
allows for matching to $\overline{\rm MS}$ without relying on large 
momentum data with their large statistical noise and potential discretization errors. 
Third,  the summation method, that allows for a better control of the excited state contamination. Having studied finite volume effects and discretization errors in~\cite{Joo:2019jct}, in our upcoming work we  plan to study  in a systematic way the continuum extrapolation and finite volume as well as effects stemming from excited state contamination and higher twist contributions.

\smallskip
\noindent\emph{ Acknowledgements}.\,---\,
 JK thanks R. Sufian for helpful comments. This work is supported by Jefferson
Science Associates, LLC under U.S. DOE Contract \#DE-AC05-06OR23177.
KO was supported in part by U.S.  DOE grant \mbox{
  \#DE-FG02-04ER41302}.
    AR was supported in part by U.S. DOE Grant
\mbox{\#DE-FG02-97ER41028. }  J.K. was supported
in part by the U.S. Department of Energy under contract
DE-FG02-04ER41302, Department of Energy Office of Science Graduate
Student Research fellowships, through the U.S. Department of Energy,
Office of Science, Office of Workforce Development for Teachers and
Scientists, Office of Science Graduate Student Research (SCGSR)
program and is supported by U.S. Department of Energy grant DE-SC0011941.  
The authors gratefully acknowledge the computing time
granted by the John von Neumann Institute for Computing (NIC) and
provided on the supercomputer JURECA at J\"ulich Supercomputing Centre
(JSC)~\cite{jureca}. We acknowledge the facilities of the USQCD Collaboration used for this research in part, which are funded by the Office of Science of the U.S. Department of Energy. This work was performed in part using computing
facilities at the College of William and Mary which were provided by
contributions from the National Science Foundation (MRI grant
PHY-1626177), and the Commonwealth of Virginia Equipment Trust Fund.
The authors acknowledge William \& Mary Research Computing for providing computational resources and/or technical support that have contributed to the results reported within this paper. This work used the Extreme Science and Engineering Discovery Environment (XSEDE), which is supported by National Science Foundation grant number ACI-1548562~\cite{xsede}.
 In addition, this work used resources at
NERSC, a DOE Office of Science User Facility supported by the Office
of Science of the U.S. Department of Energy under Contract
\#DE-AC02-05CH11231, as well as resources of the Oak Ridge Leadership Computing Facility at the Oak Ridge National Laboratory, which is supported by the Office of Science of the U.S. Department of Energy under Contract No. \mbox{\#DE-AC05-00OR22725}.  The software codes
{\tt Chroma}~\cite{Edwards:2004sx}, {\tt QUDA}~\cite{Clark:2009wm,Babich:2010mu} and {\tt QPhiX}~\cite{QPhiX2} were used. 
The authors acknowledge support from the U.S. Department of Energy, Office of Science, Office of Advanced Scientific Computing Research and Office of Nuclear Physics, Scientific Discovery through Advanced Computing (SciDAC) program, and of the U.S. Department of Energy Exascale Computing Project.

\section{Supplementary Materials}
\subsection{Correlations}
\begin{figure}[!htp]
\centering
\includegraphics[width=0.48\textwidth]{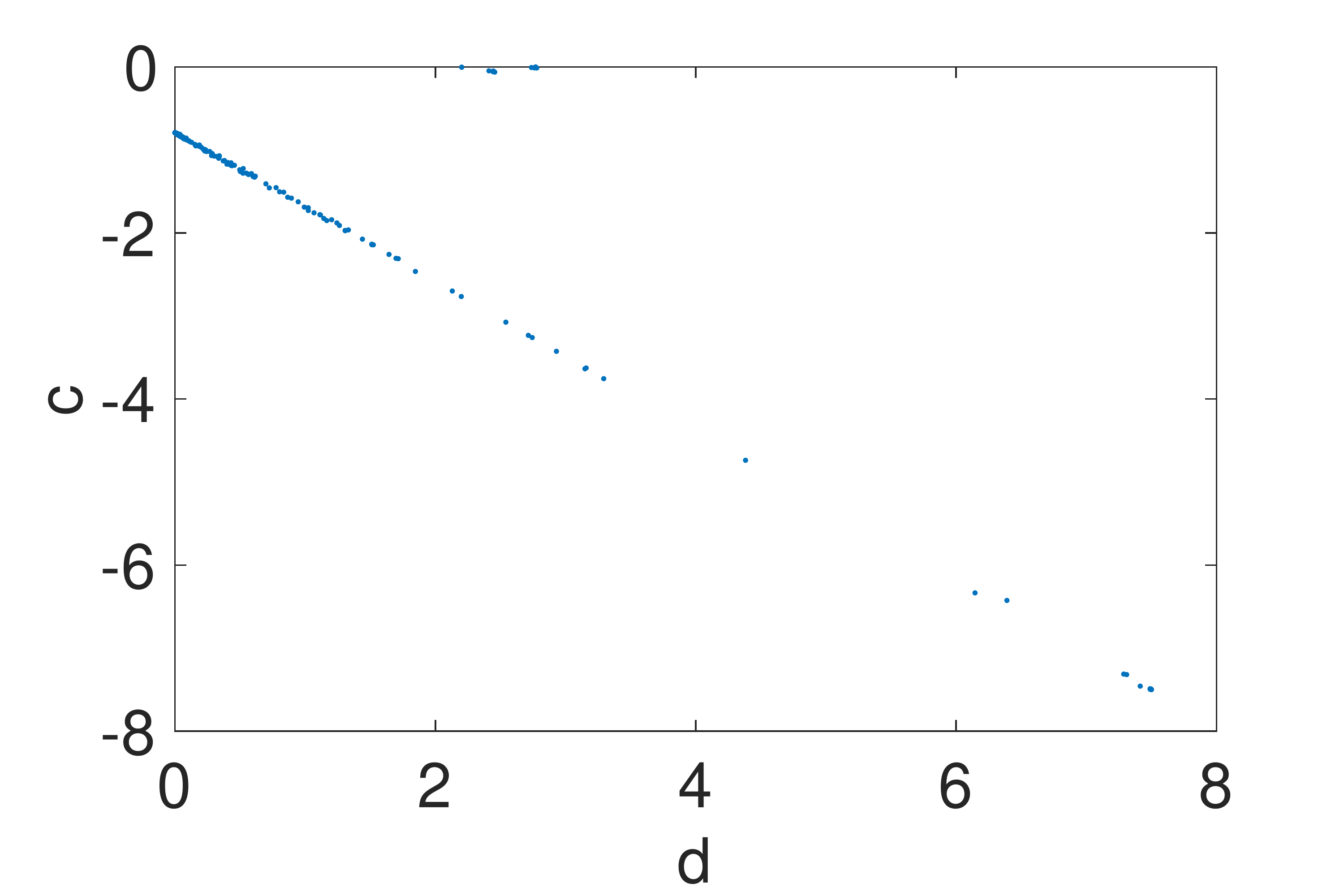}
\includegraphics[width=0.48\textwidth]{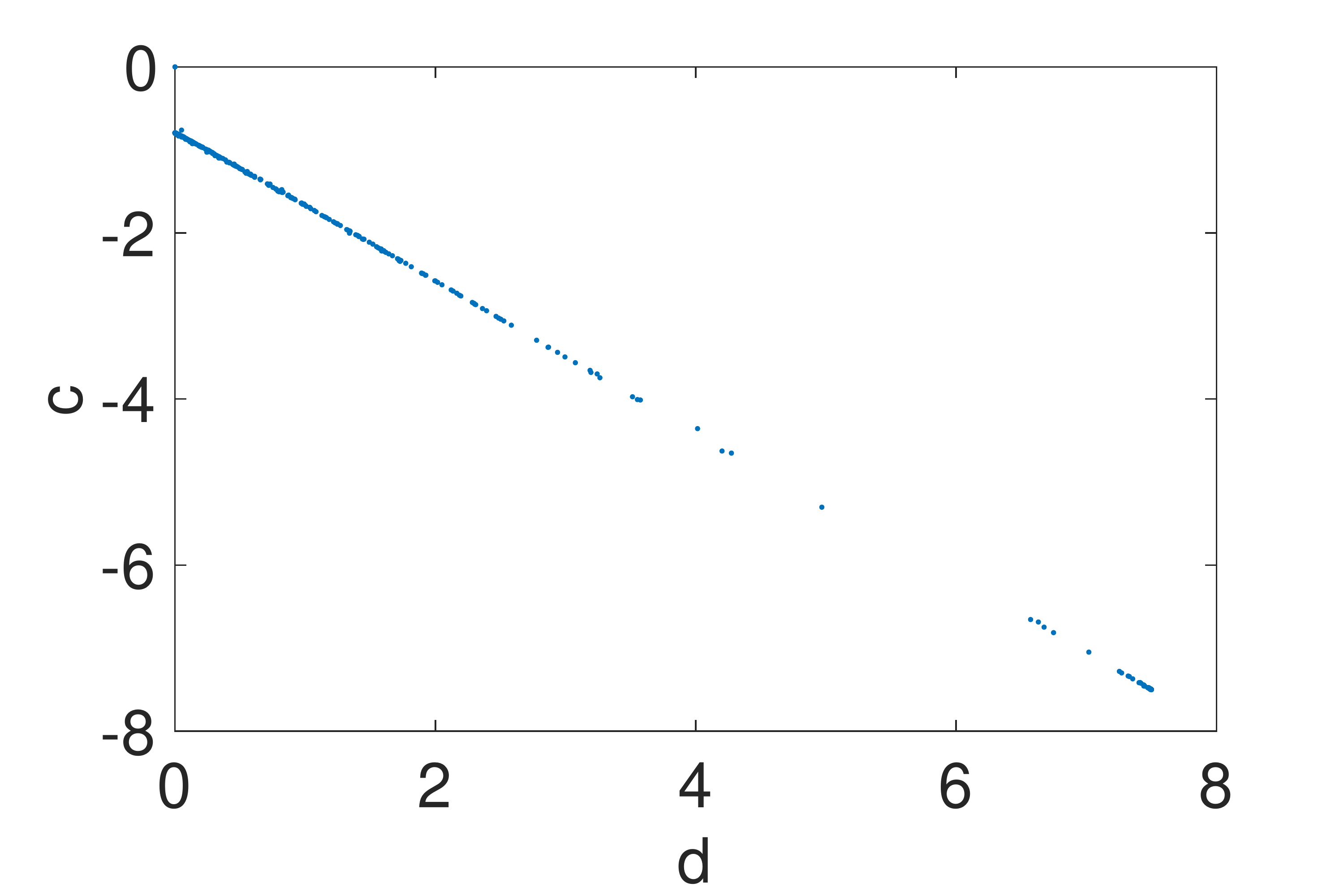}
\includegraphics[width=0.48\textwidth]{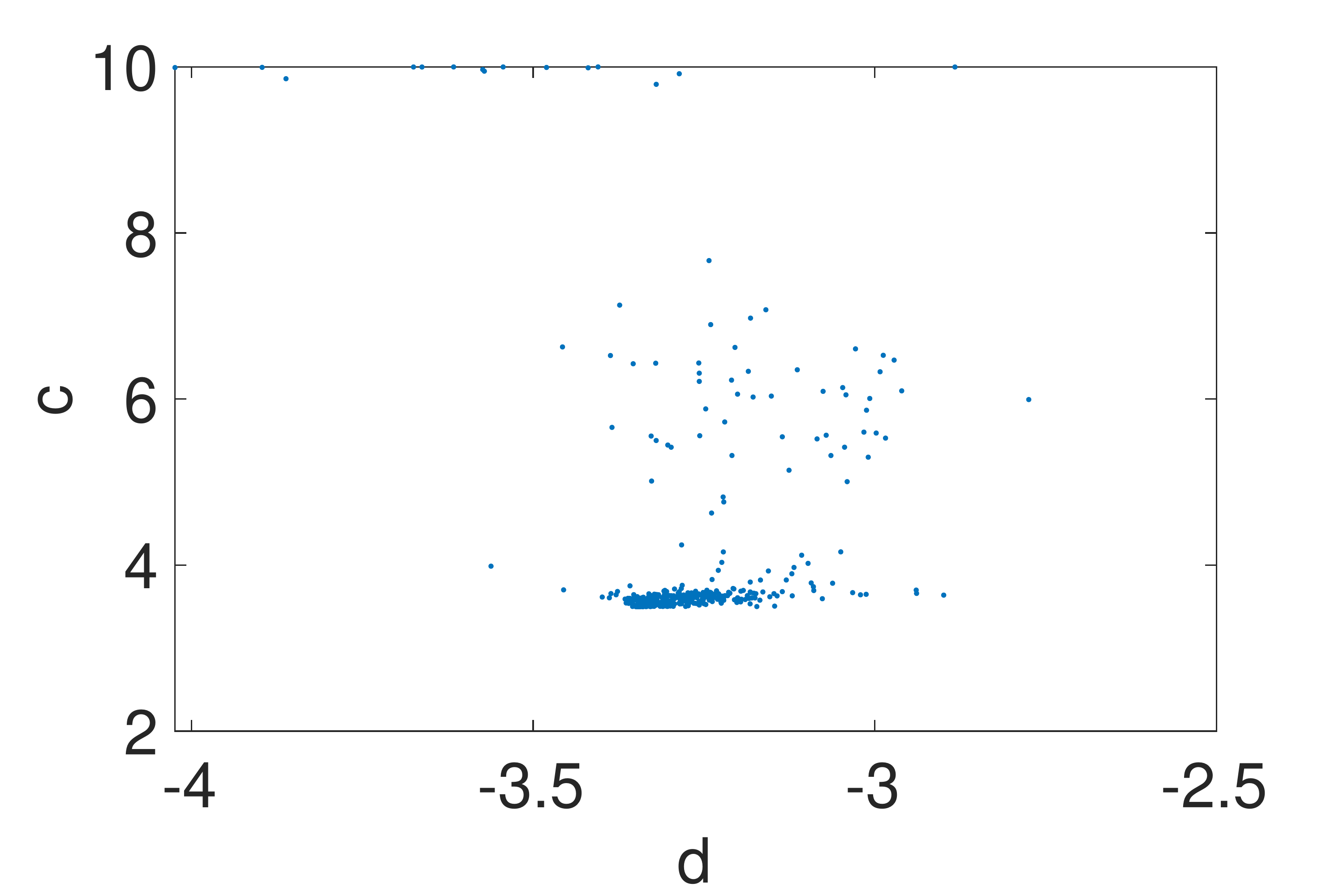}
\caption{\label{fig:correlation_c_d} The correlations between the parameters $c$ and $d$ in Eq.~\ref{eq:pdf_fit_form} for the PDF fits. The fits for the data with pion mass 172 MeV, 278 MeV, and 358 MeV are in the top, middle, and bottom panels, respectively. The lighter two pion masses display a very strong correlation between these parameters, implying a non-trivial cancellation is used to reproduce the data. }
\end{figure}

When fitting a functional form, there can exist non-trivial correlations between the resulting parameters. Evidence of this occurring with the $c$ and $d$ parameters form Eq.~\ref{eq:pdf_fit_form} appears in the results from the data with the two lighter pion masses. On the other hand, these parameters do not appear nearly as correlated in the heaviest pion mass, instead having a much larger variance. Plots of the correlations between the parameters are shown in Fig.~\ref{fig:correlation_c_d}. The lighter two pion mass results appear to have an extremely strong correlation between these two parameters, likely from a cancellation occurring to describe the data. The heaviest pion mass result's much larger variance in these parameters leads to the much larger variance in the PDF. Different functional forms can be used to study the systematic error created by these correlations. A future work will include a systematic study of many functional forms and their correlated parameters.

\subsection{Comparison with other models}
\begin{figure}[!htp]
\centering
\subfigure[]{\includegraphics[width=0.48\textwidth]{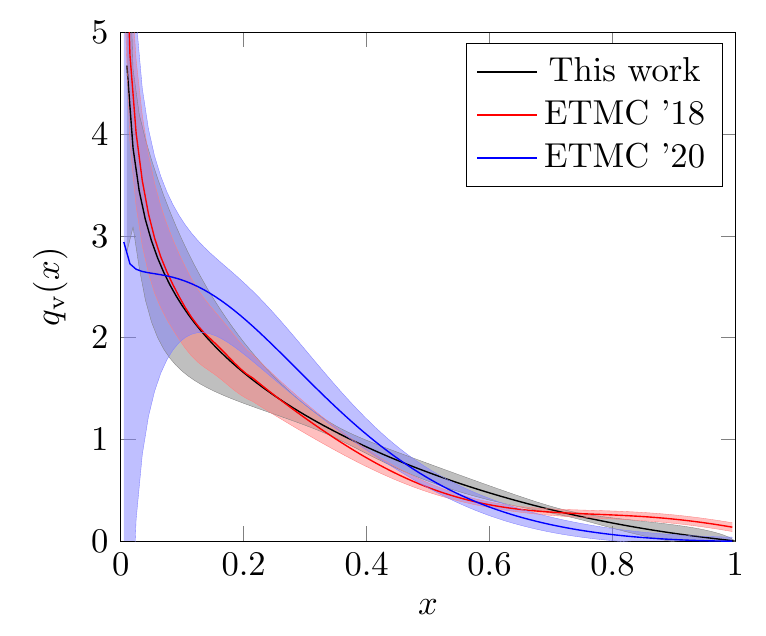}\label{etmccomp}}
\caption{\label{fig:etmccomp} Our determination of the nucleon light-cone PDF at the physical pion mass, compared to the 2018 results of ETMC employing the method of quasi-PDFs and compared to the 2020 results of ETMC which is a re-analysis of the same lattice data albeit with the method of pseudo-PDFs.}
\end{figure}

Here we would like to compare our results with other determinations of the light cone nucleon PDF at the physical point. The Extended Twisted Mass collaboration (ETMC) published in 2018 their analysis employing one ensemble of twisted mass fermions at the physical pion mass, with a lattice spacing of 0.0938 fm in a $48^3\times 96$ box, employing the method of the quasi PDFs~\cite{Alexandrou:2018pbm}. The properties of their gauge configurations are very similar to ours and this allows for a meaningful comparison. These results are labeled as ETMC '18 in 
Fig.~\ref{etmccomp}. In 2020, shortly after our preprint appeared on arXiv, they reanalyzed the same lattice data employing the method of pseudo-PDFs~\cite{Bhat:2020ktg} that has been developed by our group. These later results are labeled as ETMC '20 in Fig.~\ref{etmccomp}. 

As can be seen, the ETMC results, particularly their pseudo-PDF results, are in good agreement with our own. In the pseudo-PDF study, ETMC implemented multiple methods of solving the inverse problem suggested in~\cite{Karpie:2019eiq} and abandoned the discrete Fourier transform used in their previous quasi-PDF calculation. The discrete Fourier transform, which had been used in all calculations of the PDFs at physical pion mass prior to our study, is one of the biggest sources of pathological systematic errors in the calculations. Much better agreement with the phenomenological determinations of the PDFs are observed when fits to a functional form are used, as was done in this work. A direct comparison of the outcomes can be seen in~\cite{Bhat:2020ktg}. Based upon this comparison, we believe this systematic error, rather than a difference in factorization methodology, can sufficiently explain the discrepancy, particularly at large $x$, between the ETMC '18 quasi-PDF results and the two pseudo-PDF results. 

%%%%%%%%%%%%%%%%%%%%%%%%%%%%%%%%%%%%%%%%%%%%%%%%%%%%%%%%%
%%                              BIBLIO
%%%%%%%%%%%%%%%%%%%%%%%%%%%%%%%%%%%%%%%%%%%%%%%%%%%%%%%%%

%\bibliographystyle{unsrt} % not to be used with revtex

\bibliography{dynppdfs.bib}

\begin{thebibliography}{68}%
\makeatletter
\providecommand \@ifxundefined [1]{%
 \@ifx{#1\undefined}
}%
\providecommand \@ifnum [1]{%
 \ifnum #1\expandafter \@firstoftwo
 \else \expandafter \@secondoftwo
 \fi
}%
\providecommand \@ifx [1]{%
 \ifx #1\expandafter \@firstoftwo
 \else \expandafter \@secondoftwo
 \fi
}%
\providecommand \natexlab [1]{#1}%
\providecommand \enquote  [1]{``#1''}%
\providecommand \bibnamefont  [1]{#1}%
\providecommand \bibfnamefont [1]{#1}%
\providecommand \citenamefont [1]{#1}%
\providecommand \href@noop [0]{\@secondoftwo}%
\providecommand \href [0]{\begingroup \@sanitize@url \@href}%
\providecommand \@href[1]{\@@startlink{#1}\@@href}%
\providecommand \@@href[1]{\endgroup#1\@@endlink}%
\providecommand \@sanitize@url [0]{\catcode `\\12\catcode `\$12\catcode
  `\&12\catcode `\#12\catcode `\^12\catcode `\_12\catcode `\%12\relax}%
\providecommand \@@startlink[1]{}%
\providecommand \@@endlink[0]{}%
\providecommand \url  [0]{\begingroup\@sanitize@url \@url }%
\providecommand \@url [1]{\endgroup\@href {#1}{\urlprefix }}%
\providecommand \urlprefix  [0]{URL }%
\providecommand \Eprint [0]{\href }%
\providecommand \doibase [0]{http://dx.doi.org/}%
\providecommand \selectlanguage [0]{\@gobble}%
\providecommand \bibinfo  [0]{\@secondoftwo}%
\providecommand \bibfield  [0]{\@secondoftwo}%
\providecommand \translation [1]{[#1]}%
\providecommand \BibitemOpen [0]{}%
\providecommand \bibitemStop [0]{}%
\providecommand \bibitemNoStop [0]{.\EOS\space}%
\providecommand \EOS [0]{\spacefactor3000\relax}%
\providecommand \BibitemShut  [1]{\csname bibitem#1\endcsname}%
\let\auto@bib@innerbib\@empty
%</preamble>
\bibitem [{\citenamefont {Gao}\ \emph {et~al.}(2018)\citenamefont {Gao},
  \citenamefont {Harland-Lang},\ and\ \citenamefont {Rojo}}]{Gao:2017yyd}%
  \BibitemOpen
  \bibfield  {author} {\bibinfo {author} {\bibfnamefont {J.}~\bibnamefont
  {Gao}}, \bibinfo {author} {\bibfnamefont {L.}~\bibnamefont {Harland-Lang}}, \
  and\ \bibinfo {author} {\bibfnamefont {J.}~\bibnamefont {Rojo}},\ }\href
  {\doibase 10.1016/j.physrep.2018.03.002} {\bibfield  {journal} {\bibinfo
  {journal} {Phys. Rept.}\ }\textbf {\bibinfo {volume} {742}},\ \bibinfo
  {pages} {1} (\bibinfo {year} {2018})},\ \Eprint
  {http://arxiv.org/abs/1709.04922} {arXiv:1709.04922 [hep-ph]} \BibitemShut
  {NoStop}%
%%CITATION = ARXIV:1709.04922;%%
\bibitem [{\citenamefont {Ji}(2013)}]{Ji:2013dva}%
  \BibitemOpen
  \bibfield  {author} {\bibinfo {author} {\bibfnamefont {X.}~\bibnamefont
  {Ji}},\ }\href {\doibase 10.1103/PhysRevLett.110.262002} {\bibfield
  {journal} {\bibinfo  {journal} {Phys. Rev. Lett.}\ }\textbf {\bibinfo
  {volume} {110}},\ \bibinfo {pages} {262002} (\bibinfo {year}
  {2013})}\BibitemShut {NoStop}%
\bibitem [{\citenamefont {Lin}\ \emph {et~al.}(2015)\citenamefont {Lin},
  \citenamefont {Chen}, \citenamefont {Cohen},\ and\ \citenamefont
  {Ji}}]{Lin:2014zya}%
  \BibitemOpen
  \bibfield  {author} {\bibinfo {author} {\bibfnamefont {H.-W.}\ \bibnamefont
  {Lin}}, \bibinfo {author} {\bibfnamefont {J.-W.}\ \bibnamefont {Chen}},
  \bibinfo {author} {\bibfnamefont {S.~D.}\ \bibnamefont {Cohen}}, \ and\
  \bibinfo {author} {\bibfnamefont {X.}~\bibnamefont {Ji}},\ }\href {\doibase
  10.1103/PhysRevD.91.054510} {\bibfield  {journal} {\bibinfo  {journal} {Phys.
  Rev.}\ }\textbf {\bibinfo {volume} {D91}},\ \bibinfo {pages} {054510}
  (\bibinfo {year} {2015})},\ \Eprint {http://arxiv.org/abs/1402.1462}
  {arXiv:1402.1462 [hep-ph]} \BibitemShut {NoStop}%
%%CITATION = ARXIV:1402.1462;%%
\bibitem [{\citenamefont {Chen}\ \emph {et~al.}(2016)\citenamefont {Chen},
  \citenamefont {Cohen}, \citenamefont {Ji}, \citenamefont {Lin},\ and\
  \citenamefont {Zhang}}]{Chen:2016utp}%
  \BibitemOpen
  \bibfield  {author} {\bibinfo {author} {\bibfnamefont {J.-W.}\ \bibnamefont
  {Chen}}, \bibinfo {author} {\bibfnamefont {S.~D.}\ \bibnamefont {Cohen}},
  \bibinfo {author} {\bibfnamefont {X.}~\bibnamefont {Ji}}, \bibinfo {author}
  {\bibfnamefont {H.-W.}\ \bibnamefont {Lin}}, \ and\ \bibinfo {author}
  {\bibfnamefont {J.-H.}\ \bibnamefont {Zhang}},\ }\href {\doibase
  10.1016/j.nuclphysb.2016.07.033} {\bibfield  {journal} {\bibinfo  {journal}
  {Nucl. Phys.}\ }\textbf {\bibinfo {volume} {B911}},\ \bibinfo {pages} {246}
  (\bibinfo {year} {2016})},\ \Eprint {http://arxiv.org/abs/1603.06664}
  {arXiv:1603.06664 [hep-ph]} \BibitemShut {NoStop}%
%%CITATION = ARXIV:1603.06664;%%
\bibitem [{\citenamefont {Alexandrou}\ \emph {et~al.}(2015)\citenamefont
  {Alexandrou}, \citenamefont {Cichy}, \citenamefont {Drach}, \citenamefont
  {Garcia-Ramos}, \citenamefont {Hadjiyiannakou}, \citenamefont {Jansen},
  \citenamefont {Steffens},\ and\ \citenamefont {Wiese}}]{Alexandrou:2015rja}%
  \BibitemOpen
  \bibfield  {author} {\bibinfo {author} {\bibfnamefont {C.}~\bibnamefont
  {Alexandrou}}, \bibinfo {author} {\bibfnamefont {K.}~\bibnamefont {Cichy}},
  \bibinfo {author} {\bibfnamefont {V.}~\bibnamefont {Drach}}, \bibinfo
  {author} {\bibfnamefont {E.}~\bibnamefont {Garcia-Ramos}}, \bibinfo {author}
  {\bibfnamefont {K.}~\bibnamefont {Hadjiyiannakou}}, \bibinfo {author}
  {\bibfnamefont {K.}~\bibnamefont {Jansen}}, \bibinfo {author} {\bibfnamefont
  {F.}~\bibnamefont {Steffens}}, \ and\ \bibinfo {author} {\bibfnamefont
  {C.}~\bibnamefont {Wiese}},\ }\href {\doibase 10.1103/PhysRevD.92.014502}
  {\bibfield  {journal} {\bibinfo  {journal} {Phys. Rev.}\ }\textbf {\bibinfo
  {volume} {D92}},\ \bibinfo {pages} {014502} (\bibinfo {year} {2015})},\
  \Eprint {http://arxiv.org/abs/1504.07455} {arXiv:1504.07455 [hep-lat]}
  \BibitemShut {NoStop}%
%%CITATION = ARXIV:1504.07455;%%
\bibitem [{\citenamefont {Alexandrou}\ \emph
  {et~al.}(2017{\natexlab{a}})\citenamefont {Alexandrou}, \citenamefont
  {Cichy}, \citenamefont {Constantinou}, \citenamefont {Hadjiyiannakou},
  \citenamefont {Jansen}, \citenamefont {Steffens},\ and\ \citenamefont
  {Wiese}}]{Alexandrou:2016jqi}%
  \BibitemOpen
  \bibfield  {author} {\bibinfo {author} {\bibfnamefont {C.}~\bibnamefont
  {Alexandrou}}, \bibinfo {author} {\bibfnamefont {K.}~\bibnamefont {Cichy}},
  \bibinfo {author} {\bibfnamefont {M.}~\bibnamefont {Constantinou}}, \bibinfo
  {author} {\bibfnamefont {K.}~\bibnamefont {Hadjiyiannakou}}, \bibinfo
  {author} {\bibfnamefont {K.}~\bibnamefont {Jansen}}, \bibinfo {author}
  {\bibfnamefont {F.}~\bibnamefont {Steffens}}, \ and\ \bibinfo {author}
  {\bibfnamefont {C.}~\bibnamefont {Wiese}},\ }\href {\doibase
  10.1103/PhysRevD.96.014513} {\bibfield  {journal} {\bibinfo  {journal} {Phys.
  Rev.}\ }\textbf {\bibinfo {volume} {D96}},\ \bibinfo {pages} {014513}
  (\bibinfo {year} {2017}{\natexlab{a}})},\ \Eprint
  {http://arxiv.org/abs/1610.03689} {arXiv:1610.03689 [hep-lat]} \BibitemShut
  {NoStop}%
%%CITATION = ARXIV:1610.03689;%%
\bibitem [{\citenamefont {Monahan}\ and\ \citenamefont
  {Orginos}(2017)}]{Monahan:2016bvm}%
  \BibitemOpen
  \bibfield  {author} {\bibinfo {author} {\bibfnamefont {C.}~\bibnamefont
  {Monahan}}\ and\ \bibinfo {author} {\bibfnamefont {K.}~\bibnamefont
  {Orginos}},\ }\href {\doibase 10.1007/JHEP03(2017)116} {\bibfield  {journal}
  {\bibinfo  {journal} {JHEP}\ }\textbf {\bibinfo {volume} {03}},\ \bibinfo
  {pages} {116} (\bibinfo {year} {2017})},\ \Eprint
  {http://arxiv.org/abs/1612.01584} {arXiv:1612.01584 [hep-lat]} \BibitemShut
  {NoStop}%
%%CITATION = ARXIV:1612.01584;%%
\bibitem [{\citenamefont {Zhang}\ \emph {et~al.}(2017)\citenamefont {Zhang},
  \citenamefont {Chen}, \citenamefont {Ji}, \citenamefont {Jin},\ and\
  \citenamefont {Lin}}]{Zhang:2017bzy}%
  \BibitemOpen
  \bibfield  {author} {\bibinfo {author} {\bibfnamefont {J.-H.}\ \bibnamefont
  {Zhang}}, \bibinfo {author} {\bibfnamefont {J.-W.}\ \bibnamefont {Chen}},
  \bibinfo {author} {\bibfnamefont {X.}~\bibnamefont {Ji}}, \bibinfo {author}
  {\bibfnamefont {L.}~\bibnamefont {Jin}}, \ and\ \bibinfo {author}
  {\bibfnamefont {H.-W.}\ \bibnamefont {Lin}},\ }\href {\doibase
  10.1103/PhysRevD.95.094514} {\bibfield  {journal} {\bibinfo  {journal} {Phys.
  Rev.}\ }\textbf {\bibinfo {volume} {D95}},\ \bibinfo {pages} {094514}
  (\bibinfo {year} {2017})},\ \Eprint {http://arxiv.org/abs/1702.00008}
  {arXiv:1702.00008 [hep-lat]} \BibitemShut {NoStop}%
%%CITATION = ARXIV:1702.00008;%%
\bibitem [{\citenamefont {Alexandrou}\ \emph
  {et~al.}(2017{\natexlab{b}})\citenamefont {Alexandrou}, \citenamefont
  {Cichy}, \citenamefont {Constantinou}, \citenamefont {Hadjiyiannakou},
  \citenamefont {Jansen}, \citenamefont {Panagopoulos},\ and\ \citenamefont
  {Steffens}}]{Alexandrou:2017huk}%
  \BibitemOpen
  \bibfield  {author} {\bibinfo {author} {\bibfnamefont {C.}~\bibnamefont
  {Alexandrou}}, \bibinfo {author} {\bibfnamefont {K.}~\bibnamefont {Cichy}},
  \bibinfo {author} {\bibfnamefont {M.}~\bibnamefont {Constantinou}}, \bibinfo
  {author} {\bibfnamefont {K.}~\bibnamefont {Hadjiyiannakou}}, \bibinfo
  {author} {\bibfnamefont {K.}~\bibnamefont {Jansen}}, \bibinfo {author}
  {\bibfnamefont {H.}~\bibnamefont {Panagopoulos}}, \ and\ \bibinfo {author}
  {\bibfnamefont {F.}~\bibnamefont {Steffens}},\ }\href {\doibase
  10.1016/j.nuclphysb.2017.08.012} {\bibfield  {journal} {\bibinfo  {journal}
  {Nucl. Phys.}\ }\textbf {\bibinfo {volume} {B923}},\ \bibinfo {pages} {394}
  (\bibinfo {year} {2017}{\natexlab{b}})},\ \Eprint
  {http://arxiv.org/abs/1706.00265} {arXiv:1706.00265 [hep-lat]} \BibitemShut
  {NoStop}%
%%CITATION = ARXIV:1706.00265;%%
\bibitem [{\citenamefont {Green}\ \emph {et~al.}(2018)\citenamefont {Green},
  \citenamefont {Jansen},\ and\ \citenamefont {Steffens}}]{Green:2017xeu}%
  \BibitemOpen
  \bibfield  {author} {\bibinfo {author} {\bibfnamefont {J.}~\bibnamefont
  {Green}}, \bibinfo {author} {\bibfnamefont {K.}~\bibnamefont {Jansen}}, \
  and\ \bibinfo {author} {\bibfnamefont {F.}~\bibnamefont {Steffens}},\ }\href
  {\doibase 10.1103/PhysRevLett.121.022004} {\bibfield  {journal} {\bibinfo
  {journal} {Phys. Rev. Lett.}\ }\textbf {\bibinfo {volume} {121}},\ \bibinfo
  {pages} {022004} (\bibinfo {year} {2018})},\ \Eprint
  {http://arxiv.org/abs/1707.07152} {arXiv:1707.07152 [hep-lat]} \BibitemShut
  {NoStop}%
%%CITATION = ARXIV:1707.07152;%%
\bibitem [{\citenamefont {Stewart}\ and\ \citenamefont
  {Zhao}(2018)}]{Stewart:2017tvs}%
  \BibitemOpen
  \bibfield  {author} {\bibinfo {author} {\bibfnamefont {I.~W.}\ \bibnamefont
  {Stewart}}\ and\ \bibinfo {author} {\bibfnamefont {Y.}~\bibnamefont {Zhao}},\
  }\href {\doibase 10.1103/PhysRevD.97.054512} {\bibfield  {journal} {\bibinfo
  {journal} {Phys. Rev.}\ }\textbf {\bibinfo {volume} {D97}},\ \bibinfo {pages}
  {054512} (\bibinfo {year} {2018})},\ \Eprint
  {http://arxiv.org/abs/1709.04933} {arXiv:1709.04933 [hep-ph]} \BibitemShut
  {NoStop}%
%%CITATION = ARXIV:1709.04933;%%
\bibitem [{\citenamefont {Monahan}(2018{\natexlab{a}})}]{Monahan:2017hpu}%
  \BibitemOpen
  \bibfield  {author} {\bibinfo {author} {\bibfnamefont {C.}~\bibnamefont
  {Monahan}},\ }\href {\doibase 10.1103/PhysRevD.97.054507} {\bibfield
  {journal} {\bibinfo  {journal} {Phys. Rev.}\ }\textbf {\bibinfo {volume}
  {D97}},\ \bibinfo {pages} {054507} (\bibinfo {year} {2018}{\natexlab{a}})},\
  \Eprint {http://arxiv.org/abs/1710.04607} {arXiv:1710.04607 [hep-lat]}
  \BibitemShut {NoStop}%
%%CITATION = ARXIV:1710.04607;%%
\bibitem [{\citenamefont {Broniowski}\ and\ \citenamefont
  {Ruiz~Arriola}(2018)}]{Broniowski:2017gfp}%
  \BibitemOpen
  \bibfield  {author} {\bibinfo {author} {\bibfnamefont {W.}~\bibnamefont
  {Broniowski}}\ and\ \bibinfo {author} {\bibfnamefont {E.}~\bibnamefont
  {Ruiz~Arriola}},\ }\href {\doibase 10.1103/PhysRevD.97.034031} {\bibfield
  {journal} {\bibinfo  {journal} {Phys. Rev.}\ }\textbf {\bibinfo {volume}
  {D97}},\ \bibinfo {pages} {034031} (\bibinfo {year} {2018})},\ \Eprint
  {http://arxiv.org/abs/1711.03377} {arXiv:1711.03377 [hep-ph]} \BibitemShut
  {NoStop}%
%%CITATION = ARXIV:1711.03377;%%
\bibitem [{\citenamefont {Alexandrou}\ \emph
  {et~al.}(2018{\natexlab{a}})\citenamefont {Alexandrou}, \citenamefont
  {Cichy}, \citenamefont {Constantinou}, \citenamefont {Jansen}, \citenamefont
  {Scapellato},\ and\ \citenamefont {Steffens}}]{Alexandrou:2018pbm}%
  \BibitemOpen
  \bibfield  {author} {\bibinfo {author} {\bibfnamefont {C.}~\bibnamefont
  {Alexandrou}}, \bibinfo {author} {\bibfnamefont {K.}~\bibnamefont {Cichy}},
  \bibinfo {author} {\bibfnamefont {M.}~\bibnamefont {Constantinou}}, \bibinfo
  {author} {\bibfnamefont {K.}~\bibnamefont {Jansen}}, \bibinfo {author}
  {\bibfnamefont {A.}~\bibnamefont {Scapellato}}, \ and\ \bibinfo {author}
  {\bibfnamefont {F.}~\bibnamefont {Steffens}},\ }\href {\doibase
  10.1103/PhysRevLett.121.112001} {\bibfield  {journal} {\bibinfo  {journal}
  {Phys. Rev. Lett.}\ }\textbf {\bibinfo {volume} {121}},\ \bibinfo {pages}
  {112001} (\bibinfo {year} {2018}{\natexlab{a}})},\ \Eprint
  {http://arxiv.org/abs/1803.02685} {arXiv:1803.02685 [hep-lat]} \BibitemShut
  {NoStop}%
%%CITATION = ARXIV:1803.02685;%%
\bibitem [{\citenamefont {Chen}\ \emph {et~al.}(2018)\citenamefont {Chen},
  \citenamefont {Jin}, \citenamefont {Lin}, \citenamefont {Liu}, \citenamefont
  {Yang}, \citenamefont {Zhang},\ and\ \citenamefont {Zhao}}]{Chen:2018xof}%
  \BibitemOpen
  \bibfield  {author} {\bibinfo {author} {\bibfnamefont {J.-W.}\ \bibnamefont
  {Chen}}, \bibinfo {author} {\bibfnamefont {L.}~\bibnamefont {Jin}}, \bibinfo
  {author} {\bibfnamefont {H.-W.}\ \bibnamefont {Lin}}, \bibinfo {author}
  {\bibfnamefont {Y.-S.}\ \bibnamefont {Liu}}, \bibinfo {author} {\bibfnamefont
  {Y.-B.}\ \bibnamefont {Yang}}, \bibinfo {author} {\bibfnamefont {J.-H.}\
  \bibnamefont {Zhang}}, \ and\ \bibinfo {author} {\bibfnamefont
  {Y.}~\bibnamefont {Zhao}},\ }\href@noop {} {\  (\bibinfo {year} {2018})},\
  \Eprint {http://arxiv.org/abs/1803.04393} {arXiv:1803.04393 [hep-lat]}
  \BibitemShut {NoStop}%
%%CITATION = ARXIV:1803.04393;%%
\bibitem [{\citenamefont {Alexandrou}\ \emph
  {et~al.}(2018{\natexlab{b}})\citenamefont {Alexandrou}, \citenamefont
  {Cichy}, \citenamefont {Constantinou}, \citenamefont {Jansen}, \citenamefont
  {Scapellato},\ and\ \citenamefont {Steffens}}]{Alexandrou:2018eet}%
  \BibitemOpen
  \bibfield  {author} {\bibinfo {author} {\bibfnamefont {C.}~\bibnamefont
  {Alexandrou}}, \bibinfo {author} {\bibfnamefont {K.}~\bibnamefont {Cichy}},
  \bibinfo {author} {\bibfnamefont {M.}~\bibnamefont {Constantinou}}, \bibinfo
  {author} {\bibfnamefont {K.}~\bibnamefont {Jansen}}, \bibinfo {author}
  {\bibfnamefont {A.}~\bibnamefont {Scapellato}}, \ and\ \bibinfo {author}
  {\bibfnamefont {F.}~\bibnamefont {Steffens}},\ }\href {\doibase
  10.1103/PhysRevD.98.091503} {\bibfield  {journal} {\bibinfo  {journal} {Phys.
  Rev.}\ }\textbf {\bibinfo {volume} {D98}},\ \bibinfo {pages} {091503}
  (\bibinfo {year} {2018}{\natexlab{b}})},\ \Eprint
  {http://arxiv.org/abs/1807.00232} {arXiv:1807.00232 [hep-lat]} \BibitemShut
  {NoStop}%
%%CITATION = ARXIV:1807.00232;%%
\bibitem [{\citenamefont {Lin}\ \emph {et~al.}(2018{\natexlab{a}})\citenamefont
  {Lin}, \citenamefont {Chen}, \citenamefont {Ji}, \citenamefont {Jin},
  \citenamefont {Li}, \citenamefont {Liu}, \citenamefont {Yang}, \citenamefont
  {Zhang},\ and\ \citenamefont {Zhao}}]{Lin:2018qky}%
  \BibitemOpen
  \bibfield  {author} {\bibinfo {author} {\bibfnamefont {H.-W.}\ \bibnamefont
  {Lin}}, \bibinfo {author} {\bibfnamefont {J.-W.}\ \bibnamefont {Chen}},
  \bibinfo {author} {\bibfnamefont {X.}~\bibnamefont {Ji}}, \bibinfo {author}
  {\bibfnamefont {L.}~\bibnamefont {Jin}}, \bibinfo {author} {\bibfnamefont
  {R.}~\bibnamefont {Li}}, \bibinfo {author} {\bibfnamefont {Y.-S.}\
  \bibnamefont {Liu}}, \bibinfo {author} {\bibfnamefont {Y.-B.}\ \bibnamefont
  {Yang}}, \bibinfo {author} {\bibfnamefont {J.-H.}\ \bibnamefont {Zhang}}, \
  and\ \bibinfo {author} {\bibfnamefont {Y.}~\bibnamefont {Zhao}},\ }\href
  {\doibase 10.1103/PhysRevLett.121.242003} {\bibfield  {journal} {\bibinfo
  {journal} {Phys. Rev. Lett.}\ }\textbf {\bibinfo {volume} {121}},\ \bibinfo
  {pages} {242003} (\bibinfo {year} {2018}{\natexlab{a}})},\ \Eprint
  {http://arxiv.org/abs/1807.07431} {arXiv:1807.07431 [hep-lat]} \BibitemShut
  {NoStop}%
%%CITATION = ARXIV:1807.07431;%%
\bibitem [{\citenamefont {Fan}\ \emph {et~al.}(2018)\citenamefont {Fan},
  \citenamefont {Yang}, \citenamefont {Anthony}, \citenamefont {Lin},\ and\
  \citenamefont {Liu}}]{Fan:2018dxu}%
  \BibitemOpen
  \bibfield  {author} {\bibinfo {author} {\bibfnamefont {Z.-Y.}\ \bibnamefont
  {Fan}}, \bibinfo {author} {\bibfnamefont {Y.-B.}\ \bibnamefont {Yang}},
  \bibinfo {author} {\bibfnamefont {A.}~\bibnamefont {Anthony}}, \bibinfo
  {author} {\bibfnamefont {H.-W.}\ \bibnamefont {Lin}}, \ and\ \bibinfo
  {author} {\bibfnamefont {K.-F.}\ \bibnamefont {Liu}},\ }\href {\doibase
  10.1103/PhysRevLett.121.242001} {\bibfield  {journal} {\bibinfo  {journal}
  {Phys. Rev. Lett.}\ }\textbf {\bibinfo {volume} {121}},\ \bibinfo {pages}
  {242001} (\bibinfo {year} {2018})},\ \Eprint
  {http://arxiv.org/abs/1808.02077} {arXiv:1808.02077 [hep-lat]} \BibitemShut
  {NoStop}%
%%CITATION = ARXIV:1808.02077;%%
\bibitem [{\citenamefont {Liu}\ \emph {et~al.}(2018)\citenamefont {Liu},
  \citenamefont {Chen}, \citenamefont {Jin}, \citenamefont {Li}, \citenamefont
  {Lin}, \citenamefont {Yang}, \citenamefont {Zhang},\ and\ \citenamefont
  {Zhao}}]{Liu:2018hxv}%
  \BibitemOpen
  \bibfield  {author} {\bibinfo {author} {\bibfnamefont {Y.-S.}\ \bibnamefont
  {Liu}}, \bibinfo {author} {\bibfnamefont {J.-W.}\ \bibnamefont {Chen}},
  \bibinfo {author} {\bibfnamefont {L.}~\bibnamefont {Jin}}, \bibinfo {author}
  {\bibfnamefont {R.}~\bibnamefont {Li}}, \bibinfo {author} {\bibfnamefont
  {H.-W.}\ \bibnamefont {Lin}}, \bibinfo {author} {\bibfnamefont {Y.-B.}\
  \bibnamefont {Yang}}, \bibinfo {author} {\bibfnamefont {J.-H.}\ \bibnamefont
  {Zhang}}, \ and\ \bibinfo {author} {\bibfnamefont {Y.}~\bibnamefont {Zhao}},\
  }\href@noop {} {\  (\bibinfo {year} {2018})},\ \Eprint
  {http://arxiv.org/abs/1810.05043} {arXiv:1810.05043 [hep-lat]} \BibitemShut
  {NoStop}%
%%CITATION = ARXIV:1810.05043;%%
\bibitem [{\citenamefont {Alexandrou}\ \emph {et~al.}(2019)\citenamefont
  {Alexandrou}, \citenamefont {Cichy}, \citenamefont {Constantinou},
  \citenamefont {Hadjiyiannakou}, \citenamefont {Jansen}, \citenamefont
  {Scapellato},\ and\ \citenamefont {Steffens}}]{Alexandrou:2019lfo}%
  \BibitemOpen
  \bibfield  {author} {\bibinfo {author} {\bibfnamefont {C.}~\bibnamefont
  {Alexandrou}}, \bibinfo {author} {\bibfnamefont {K.}~\bibnamefont {Cichy}},
  \bibinfo {author} {\bibfnamefont {M.}~\bibnamefont {Constantinou}}, \bibinfo
  {author} {\bibfnamefont {K.}~\bibnamefont {Hadjiyiannakou}}, \bibinfo
  {author} {\bibfnamefont {K.}~\bibnamefont {Jansen}}, \bibinfo {author}
  {\bibfnamefont {A.}~\bibnamefont {Scapellato}}, \ and\ \bibinfo {author}
  {\bibfnamefont {F.}~\bibnamefont {Steffens}},\ }\href {\doibase
  10.1103/PhysRevD.99.114504} {\bibfield  {journal} {\bibinfo  {journal} {Phys.
  Rev.}\ }\textbf {\bibinfo {volume} {D99}},\ \bibinfo {pages} {114504}
  (\bibinfo {year} {2019})},\ \Eprint {http://arxiv.org/abs/1902.00587}
  {arXiv:1902.00587 [hep-lat]} \BibitemShut {NoStop}%
%%CITATION = ARXIV:1902.00587;%%
\bibitem [{\citenamefont {Izubuchi}\ \emph {et~al.}(2019)\citenamefont
  {Izubuchi}, \citenamefont {Jin}, \citenamefont {Kallidonis}, \citenamefont
  {Karthik}, \citenamefont {Mukherjee}, \citenamefont {Petreczky},
  \citenamefont {Shugert},\ and\ \citenamefont {Syritsyn}}]{Izubuchi:2019lyk}%
  \BibitemOpen
  \bibfield  {author} {\bibinfo {author} {\bibfnamefont {T.}~\bibnamefont
  {Izubuchi}}, \bibinfo {author} {\bibfnamefont {L.}~\bibnamefont {Jin}},
  \bibinfo {author} {\bibfnamefont {C.}~\bibnamefont {Kallidonis}}, \bibinfo
  {author} {\bibfnamefont {N.}~\bibnamefont {Karthik}}, \bibinfo {author}
  {\bibfnamefont {S.}~\bibnamefont {Mukherjee}}, \bibinfo {author}
  {\bibfnamefont {P.}~\bibnamefont {Petreczky}}, \bibinfo {author}
  {\bibfnamefont {C.}~\bibnamefont {Shugert}}, \ and\ \bibinfo {author}
  {\bibfnamefont {S.}~\bibnamefont {Syritsyn}},\ }\href {\doibase
  10.1103/PhysRevD.100.034516} {\bibfield  {journal} {\bibinfo  {journal}
  {Phys. Rev.}\ }\textbf {\bibinfo {volume} {D100}},\ \bibinfo {pages} {034516}
  (\bibinfo {year} {2019})},\ \Eprint {http://arxiv.org/abs/1905.06349}
  {arXiv:1905.06349 [hep-lat]} \BibitemShut {NoStop}%
%%CITATION = ARXIV:1905.06349;%%
\bibitem [{\citenamefont {Green}\ \emph {et~al.}(2020)\citenamefont {Green},
  \citenamefont {Jansen},\ and\ \citenamefont {Steffens}}]{Green:2020xco}%
  \BibitemOpen
  \bibfield  {author} {\bibinfo {author} {\bibfnamefont {J.~R.}\ \bibnamefont
  {Green}}, \bibinfo {author} {\bibfnamefont {K.}~\bibnamefont {Jansen}}, \
  and\ \bibinfo {author} {\bibfnamefont {F.}~\bibnamefont {Steffens}},\
  }\href@noop {} {\  (\bibinfo {year} {2020})},\ \Eprint
  {http://arxiv.org/abs/2002.09408} {arXiv:2002.09408 [hep-lat]} \BibitemShut
  {NoStop}%
%%CITATION = ARXIV:2002.09408;%%
\bibitem [{\citenamefont {Chai}\ \emph {et~al.}(2020)\citenamefont {Chai} \emph
  {et~al.}}]{Chai:2020nxw}%
  \BibitemOpen
  \bibfield  {author} {\bibinfo {author} {\bibfnamefont {Y.}~\bibnamefont
  {Chai}} \emph {et~al.},\ }\href@noop {} {\  (\bibinfo {year} {2020})},\
  \Eprint {http://arxiv.org/abs/2002.12044} {arXiv:2002.12044 [hep-lat]}
  \BibitemShut {NoStop}%
%%CITATION = ARXIV:2002.12044;%%
\bibitem [{\citenamefont {Lin}\ \emph {et~al.}(2020)\citenamefont {Lin},
  \citenamefont {Chen}, \citenamefont {Fan}, \citenamefont {Zhang},\ and\
  \citenamefont {Zhang}}]{Lin:2020ssv}%
  \BibitemOpen
  \bibfield  {author} {\bibinfo {author} {\bibfnamefont {H.-W.}\ \bibnamefont
  {Lin}}, \bibinfo {author} {\bibfnamefont {J.-W.}\ \bibnamefont {Chen}},
  \bibinfo {author} {\bibfnamefont {Z.}~\bibnamefont {Fan}}, \bibinfo {author}
  {\bibfnamefont {J.-H.}\ \bibnamefont {Zhang}}, \ and\ \bibinfo {author}
  {\bibfnamefont {R.}~\bibnamefont {Zhang}},\ }\href@noop {} {\  (\bibinfo
  {year} {2020})},\ \Eprint {http://arxiv.org/abs/2003.14128} {arXiv:2003.14128
  [hep-lat]} \BibitemShut {NoStop}%
%%CITATION = ARXIV:2003.14128;%%
\bibitem [{\citenamefont {Detmold}\ and\ \citenamefont
  {Lin}(2006)}]{Detmold:2005gg}%
  \BibitemOpen
  \bibfield  {author} {\bibinfo {author} {\bibfnamefont {W.}~\bibnamefont
  {Detmold}}\ and\ \bibinfo {author} {\bibfnamefont {C.~J.~D.}\ \bibnamefont
  {Lin}},\ }\href {\doibase 10.1103/PhysRevD.73.014501} {\bibfield  {journal}
  {\bibinfo  {journal} {Phys. Rev.}\ }\textbf {\bibinfo {volume} {D73}},\
  \bibinfo {pages} {014501} (\bibinfo {year} {2006})},\ \Eprint
  {http://arxiv.org/abs/hep-lat/0507007} {arXiv:hep-lat/0507007 [hep-lat]}
  \BibitemShut {NoStop}%
%%CITATION = HEP-LAT/0507007;%%
\bibitem [{\citenamefont {Braun}\ and\ \citenamefont
  {M{\"u}ller}(2008)}]{Braun:2007wv}%
  \BibitemOpen
  \bibfield  {author} {\bibinfo {author} {\bibfnamefont {V.}~\bibnamefont
  {Braun}}\ and\ \bibinfo {author} {\bibfnamefont {D.}~\bibnamefont
  {M{\"u}ller}},\ }\href {\doibase 10.1140/epjc/s10052-008-0608-4} {\bibfield
  {journal} {\bibinfo  {journal} {Eur. Phys. J.}\ }\textbf {\bibinfo {volume}
  {C55}},\ \bibinfo {pages} {349} (\bibinfo {year} {2008})},\ \Eprint
  {http://arxiv.org/abs/0709.1348} {arXiv:0709.1348 [hep-ph]} \BibitemShut
  {NoStop}%
%%CITATION = ARXIV:0709.1348;%%
\bibitem [{\citenamefont {Chambers}\ \emph {et~al.}(2017)\citenamefont
  {Chambers}, \citenamefont {Horsley}, \citenamefont {Nakamura}, \citenamefont
  {Perlt}, \citenamefont {Rakow}, \citenamefont {Schierholz}, \citenamefont
  {Schiller}, \citenamefont {Somfleth}, \citenamefont {Young},\ and\
  \citenamefont {Zanotti}}]{Chambers:2017dov}%
  \BibitemOpen
  \bibfield  {author} {\bibinfo {author} {\bibfnamefont {A.~J.}\ \bibnamefont
  {Chambers}}, \bibinfo {author} {\bibfnamefont {R.}~\bibnamefont {Horsley}},
  \bibinfo {author} {\bibfnamefont {Y.}~\bibnamefont {Nakamura}}, \bibinfo
  {author} {\bibfnamefont {H.}~\bibnamefont {Perlt}}, \bibinfo {author}
  {\bibfnamefont {P.~E.~L.}\ \bibnamefont {Rakow}}, \bibinfo {author}
  {\bibfnamefont {G.}~\bibnamefont {Schierholz}}, \bibinfo {author}
  {\bibfnamefont {A.}~\bibnamefont {Schiller}}, \bibinfo {author}
  {\bibfnamefont {K.}~\bibnamefont {Somfleth}}, \bibinfo {author}
  {\bibfnamefont {R.~D.}\ \bibnamefont {Young}}, \ and\ \bibinfo {author}
  {\bibfnamefont {J.~M.}\ \bibnamefont {Zanotti}},\ }\href {\doibase
  10.1103/PhysRevLett.118.242001} {\bibfield  {journal} {\bibinfo  {journal}
  {Phys. Rev. Lett.}\ }\textbf {\bibinfo {volume} {118}},\ \bibinfo {pages}
  {242001} (\bibinfo {year} {2017})},\ \Eprint
  {http://arxiv.org/abs/1703.01153} {arXiv:1703.01153 [hep-lat]} \BibitemShut
  {NoStop}%
%%CITATION = ARXIV:1703.01153;%%
\bibitem [{\citenamefont {Liang}\ \emph {et~al.}(2019)\citenamefont {Liang},
  \citenamefont {Draper}, \citenamefont {Liu}, \citenamefont {Rothkopf},\ and\
  \citenamefont {Yang}}]{Liang:2019frk}%
  \BibitemOpen
  \bibfield  {author} {\bibinfo {author} {\bibfnamefont {J.}~\bibnamefont
  {Liang}}, \bibinfo {author} {\bibfnamefont {T.}~\bibnamefont {Draper}},
  \bibinfo {author} {\bibfnamefont {K.-F.}\ \bibnamefont {Liu}}, \bibinfo
  {author} {\bibfnamefont {A.}~\bibnamefont {Rothkopf}}, \ and\ \bibinfo
  {author} {\bibfnamefont {Y.-B.}\ \bibnamefont {Yang}},\ }\href@noop {} {\
  (\bibinfo {year} {2019})},\ \Eprint {http://arxiv.org/abs/1906.05312}
  {arXiv:1906.05312 [hep-ph]} \BibitemShut {NoStop}%
%%CITATION = ARXIV:1906.05312;%%
\bibitem [{\citenamefont {Ma}\ and\ \citenamefont {Qiu}(2018)}]{Ma:2017pxb}%
  \BibitemOpen
  \bibfield  {author} {\bibinfo {author} {\bibfnamefont {Y.-Q.}\ \bibnamefont
  {Ma}}\ and\ \bibinfo {author} {\bibfnamefont {J.-W.}\ \bibnamefont {Qiu}},\
  }\href {\doibase 10.1103/PhysRevLett.120.022003} {\bibfield  {journal}
  {\bibinfo  {journal} {Phys. Rev. Lett.}\ }\textbf {\bibinfo {volume} {120}},\
  \bibinfo {pages} {022003} (\bibinfo {year} {2018})},\ \Eprint
  {http://arxiv.org/abs/1709.03018} {arXiv:1709.03018 [hep-ph]} \BibitemShut
  {NoStop}%
%%CITATION = ARXIV:1709.03018;%%
\bibitem [{\citenamefont {Bali}\ \emph
  {et~al.}(2018{\natexlab{a}})\citenamefont {Bali} \emph
  {et~al.}}]{Bali:2017gfr}%
  \BibitemOpen
  \bibfield  {author} {\bibinfo {author} {\bibfnamefont {G.~S.}\ \bibnamefont
  {Bali}} \emph {et~al.},\ }\bibfield  {booktitle} {\emph {\bibinfo {booktitle}
  {{Proceedings, 35th International Symposium on Lattice Field Theory (Lattice
  2017): Granada, Spain, June 18-24, 2017}}},\ }\href {\doibase
  10.1140/epjc/s10052-018-5700-9} {\bibfield  {journal} {\bibinfo  {journal}
  {Eur. Phys. J.}\ }\textbf {\bibinfo {volume} {C78}},\ \bibinfo {pages} {217}
  (\bibinfo {year} {2018}{\natexlab{a}})},\ \Eprint
  {http://arxiv.org/abs/1709.04325} {arXiv:1709.04325 [hep-lat]} \BibitemShut
  {NoStop}%
%%CITATION = ARXIV:1709.04325;%%
\bibitem [{\citenamefont {Bali}\ \emph
  {et~al.}(2018{\natexlab{b}})\citenamefont {Bali}, \citenamefont {Braun},
  \citenamefont {Gläßle}, \citenamefont {G{\"o}ckeler}, \citenamefont
  {Gruber}, \citenamefont {Hutzler}, \citenamefont {Korcyl}, \citenamefont
  {Sch{\"a}fer}, \citenamefont {Wein},\ and\ \citenamefont
  {Zhang}}]{Bali:2018spj}%
  \BibitemOpen
  \bibfield  {author} {\bibinfo {author} {\bibfnamefont {G.~S.}\ \bibnamefont
  {Bali}}, \bibinfo {author} {\bibfnamefont {V.~M.}\ \bibnamefont {Braun}},
  \bibinfo {author} {\bibfnamefont {B.}~\bibnamefont {Gläßle}}, \bibinfo
  {author} {\bibfnamefont {M.}~\bibnamefont {G{\"o}ckeler}}, \bibinfo {author}
  {\bibfnamefont {M.}~\bibnamefont {Gruber}}, \bibinfo {author} {\bibfnamefont
  {F.}~\bibnamefont {Hutzler}}, \bibinfo {author} {\bibfnamefont
  {P.}~\bibnamefont {Korcyl}}, \bibinfo {author} {\bibfnamefont
  {A.}~\bibnamefont {Sch{\"a}fer}}, \bibinfo {author} {\bibfnamefont
  {P.}~\bibnamefont {Wein}}, \ and\ \bibinfo {author} {\bibfnamefont {J.-H.}\
  \bibnamefont {Zhang}},\ }\href {\doibase 10.1103/PhysRevD.98.094507}
  {\bibfield  {journal} {\bibinfo  {journal} {Phys. Rev.}\ }\textbf {\bibinfo
  {volume} {D98}},\ \bibinfo {pages} {094507} (\bibinfo {year}
  {2018}{\natexlab{b}})},\ \Eprint {http://arxiv.org/abs/1807.06671}
  {arXiv:1807.06671 [hep-lat]} \BibitemShut {NoStop}%
%%CITATION = ARXIV:1807.06671;%%
\bibitem [{\citenamefont {Sufian}\ \emph {et~al.}(2019)\citenamefont {Sufian},
  \citenamefont {Karpie}, \citenamefont {Egerer}, \citenamefont {Orginos},
  \citenamefont {Qiu},\ and\ \citenamefont {Richards}}]{Sufian:2019bol}%
  \BibitemOpen
  \bibfield  {author} {\bibinfo {author} {\bibfnamefont {R.~S.}\ \bibnamefont
  {Sufian}}, \bibinfo {author} {\bibfnamefont {J.}~\bibnamefont {Karpie}},
  \bibinfo {author} {\bibfnamefont {C.}~\bibnamefont {Egerer}}, \bibinfo
  {author} {\bibfnamefont {K.}~\bibnamefont {Orginos}}, \bibinfo {author}
  {\bibfnamefont {J.-W.}\ \bibnamefont {Qiu}}, \ and\ \bibinfo {author}
  {\bibfnamefont {D.~G.}\ \bibnamefont {Richards}},\ }\href {\doibase
  10.1103/PhysRevD.99.074507} {\bibfield  {journal} {\bibinfo  {journal} {Phys.
  Rev.}\ }\textbf {\bibinfo {volume} {D99}},\ \bibinfo {pages} {074507}
  (\bibinfo {year} {2019})},\ \Eprint {http://arxiv.org/abs/1901.03921}
  {arXiv:1901.03921 [hep-lat]} \BibitemShut {NoStop}%
%%CITATION = ARXIV:1901.03921;%%
\bibitem [{\citenamefont {Bali}\ \emph {et~al.}(2019)\citenamefont {Bali} \emph
  {et~al.}}]{Bali:2019ecy}%
  \BibitemOpen
  \bibfield  {author} {\bibinfo {author} {\bibfnamefont {G.~S.}\ \bibnamefont
  {Bali}} \emph {et~al.} (\bibinfo {collaboration} {RQCD}),\ }\href {\doibase
  10.1140/epja/i2019-12803-6} {\bibfield  {journal} {\bibinfo  {journal} {Eur.
  Phys. J.}\ }\textbf {\bibinfo {volume} {A55}},\ \bibinfo {pages} {116}
  (\bibinfo {year} {2019})},\ \Eprint {http://arxiv.org/abs/1903.12590}
  {arXiv:1903.12590 [hep-lat]} \BibitemShut {NoStop}%
%%CITATION = ARXIV:1903.12590;%%
\bibitem [{\citenamefont {Sufian}\ \emph {et~al.}(2020)\citenamefont {Sufian},
  \citenamefont {Egerer}, \citenamefont {Karpie}, \citenamefont {Edwards},
  \citenamefont {Jo\'o}, \citenamefont {Ma}, \citenamefont {Orginos},
  \citenamefont {Qiu},\ and\ \citenamefont {Richards}}]{Sufian:2020vzb}%
  \BibitemOpen
  \bibfield  {author} {\bibinfo {author} {\bibfnamefont {R.~S.}\ \bibnamefont
  {Sufian}}, \bibinfo {author} {\bibfnamefont {C.}~\bibnamefont {Egerer}},
  \bibinfo {author} {\bibfnamefont {J.}~\bibnamefont {Karpie}}, \bibinfo
  {author} {\bibfnamefont {R.~G.}\ \bibnamefont {Edwards}}, \bibinfo {author}
  {\bibfnamefont {B.}~\bibnamefont {Jo\'o}}, \bibinfo {author} {\bibfnamefont
  {Y.-Q.}\ \bibnamefont {Ma}}, \bibinfo {author} {\bibfnamefont
  {K.}~\bibnamefont {Orginos}}, \bibinfo {author} {\bibfnamefont {J.-W.}\
  \bibnamefont {Qiu}}, \ and\ \bibinfo {author} {\bibfnamefont {D.~G.}\
  \bibnamefont {Richards}},\ }\href@noop {} {\  (\bibinfo {year} {2020})},\
  \Eprint {http://arxiv.org/abs/2001.04960} {arXiv:2001.04960 [hep-lat]}
  \BibitemShut {NoStop}%
%%CITATION = ARXIV:2001.04960;%%
\bibitem [{\citenamefont {Lin}\ \emph {et~al.}(2018{\natexlab{b}})\citenamefont
  {Lin} \emph {et~al.}}]{Lin:2017snn}%
  \BibitemOpen
  \bibfield  {author} {\bibinfo {author} {\bibfnamefont {H.-W.}\ \bibnamefont
  {Lin}} \emph {et~al.},\ }\href {\doibase 10.1016/j.ppnp.2018.01.007}
  {\bibfield  {journal} {\bibinfo  {journal} {Prog. Part. Nucl. Phys.}\
  }\textbf {\bibinfo {volume} {100}},\ \bibinfo {pages} {107} (\bibinfo {year}
  {2018}{\natexlab{b}})},\ \Eprint {http://arxiv.org/abs/1711.07916}
  {arXiv:1711.07916 [hep-ph]} \BibitemShut {NoStop}%
%%CITATION = ARXIV:1711.07916;%%
\bibitem [{\citenamefont {Cichy}\ and\ \citenamefont
  {Constantinou}(2019)}]{Cichy:2018mum}%
  \BibitemOpen
  \bibfield  {author} {\bibinfo {author} {\bibfnamefont {K.}~\bibnamefont
  {Cichy}}\ and\ \bibinfo {author} {\bibfnamefont {M.}~\bibnamefont
  {Constantinou}},\ }\href {\doibase 10.1155/2019/3036904} {\bibfield
  {journal} {\bibinfo  {journal} {Adv. High Energy Phys.}\ }\textbf {\bibinfo
  {volume} {2019}},\ \bibinfo {pages} {3036904} (\bibinfo {year} {2019})},\
  \Eprint {http://arxiv.org/abs/1811.07248} {arXiv:1811.07248 [hep-lat]}
  \BibitemShut {NoStop}%
%%CITATION = ARXIV:1811.07248;%%
\bibitem [{\citenamefont {Monahan}(2018{\natexlab{b}})}]{Monahan:2018euv}%
  \BibitemOpen
  \bibfield  {author} {\bibinfo {author} {\bibfnamefont {C.}~\bibnamefont
  {Monahan}},\ }\bibfield  {booktitle} {\emph {\bibinfo {booktitle} {{36th
  International Symposium on Lattice Field Theory (Lattice 2018) East Lansing,
  MI, United States, July 22-28, 2018}}},\ }\href@noop {} {\bibfield  {journal}
  {\bibinfo  {journal} {PoS}\ }\textbf {\bibinfo {volume} {LATTICE2018}},\
  \bibinfo {pages} {018} (\bibinfo {year} {2018}{\natexlab{b}})},\ \Eprint
  {http://arxiv.org/abs/1811.00678} {arXiv:1811.00678 [hep-lat]} \BibitemShut
  {NoStop}%
%%CITATION = ARXIV:1811.00678;%%
\bibitem [{\citenamefont {Qiu}(2019)}]{Qiu:2019kyy}%
  \BibitemOpen
  \bibfield  {author} {\bibinfo {author} {\bibfnamefont {J.-W.}\ \bibnamefont
  {Qiu}},\ }in\ \href@noop {} {\emph {\bibinfo {booktitle} {{8th International
  Conference on Quarks and Nuclear Physics (QNP2018) Tsukuba, Japan, November
  13-17, 2018}}}}\ (\bibinfo {year} {2019})\ \Eprint
  {http://arxiv.org/abs/1903.11902} {arXiv:1903.11902 [hep-ph]} \BibitemShut
  {NoStop}%
%%CITATION = ARXIV:1903.11902;%%
\bibitem [{\citenamefont {Radyushkin}(2017)}]{Radyushkin:2017cyf}%
  \BibitemOpen
  \bibfield  {author} {\bibinfo {author} {\bibfnamefont {A.~V.}\ \bibnamefont
  {Radyushkin}},\ }\href {\doibase 10.1103/PhysRevD.96.034025} {\bibfield
  {journal} {\bibinfo  {journal} {Phys. Rev.}\ }\textbf {\bibinfo {volume}
  {D96}},\ \bibinfo {pages} {034025} (\bibinfo {year} {2017})},\ \Eprint
  {http://arxiv.org/abs/1705.01488} {arXiv:1705.01488 [hep-ph]} \BibitemShut
  {NoStop}%
%%CITATION = ARXIV:1705.01488;%%
\bibitem [{\citenamefont {Ioffe}(1969)}]{Ioffe:1969kf}%
  \BibitemOpen
  \bibfield  {author} {\bibinfo {author} {\bibfnamefont {B.~L.}\ \bibnamefont
  {Ioffe}},\ }\href {\doibase 10.1016/0370-2693(69)90415-8} {\bibfield
  {journal} {\bibinfo  {journal} {Phys. Lett.}\ }\textbf {\bibinfo {volume}
  {30B}},\ \bibinfo {pages} {123} (\bibinfo {year} {1969})}\BibitemShut
  {NoStop}%
%%CITATION = PHLTA,30B,123;%%
\bibitem [{\citenamefont {Braun}\ \emph {et~al.}(1995)\citenamefont {Braun},
  \citenamefont {Gornicki},\ and\ \citenamefont {Mankiewicz}}]{Braun:1994jq}%
  \BibitemOpen
  \bibfield  {author} {\bibinfo {author} {\bibfnamefont {V.}~\bibnamefont
  {Braun}}, \bibinfo {author} {\bibfnamefont {P.}~\bibnamefont {Gornicki}}, \
  and\ \bibinfo {author} {\bibfnamefont {L.}~\bibnamefont {Mankiewicz}},\
  }\href {\doibase 10.1103/PhysRevD.51.6036} {\bibfield  {journal} {\bibinfo
  {journal} {Phys. Rev.}\ }\textbf {\bibinfo {volume} {D51}},\ \bibinfo {pages}
  {6036} (\bibinfo {year} {1995})},\ \Eprint
  {http://arxiv.org/abs/hep-ph/9410318} {arXiv:hep-ph/9410318 [hep-ph]}
  \BibitemShut {NoStop}%
%%CITATION = HEP-PH/9410318;%%
\bibitem [{\citenamefont {Orginos}\ \emph {et~al.}(2017)\citenamefont
  {Orginos}, \citenamefont {Radyushkin}, \citenamefont {Karpie},\ and\
  \citenamefont {Zafeiropoulos}}]{Orginos:2017kos}%
  \BibitemOpen
  \bibfield  {author} {\bibinfo {author} {\bibfnamefont {K.}~\bibnamefont
  {Orginos}}, \bibinfo {author} {\bibfnamefont {A.}~\bibnamefont {Radyushkin}},
  \bibinfo {author} {\bibfnamefont {J.}~\bibnamefont {Karpie}}, \ and\ \bibinfo
  {author} {\bibfnamefont {S.}~\bibnamefont {Zafeiropoulos}},\ }\href {\doibase
  10.1103/PhysRevD.96.094503} {\bibfield  {journal} {\bibinfo  {journal} {Phys.
  Rev.}\ }\textbf {\bibinfo {volume} {D96}},\ \bibinfo {pages} {094503}
  (\bibinfo {year} {2017})},\ \Eprint {http://arxiv.org/abs/1706.05373}
  {arXiv:1706.05373 [hep-ph]} \BibitemShut {NoStop}%
%%CITATION = ARXIV:1706.05373;%%
\bibitem [{\citenamefont {Karpie}\ \emph
  {et~al.}(2018{\natexlab{a}})\citenamefont {Karpie}, \citenamefont {Orginos},
  \citenamefont {Radyushkin},\ and\ \citenamefont
  {Zafeiropoulos}}]{Karpie:2017bzm}%
  \BibitemOpen
  \bibfield  {author} {\bibinfo {author} {\bibfnamefont {J.}~\bibnamefont
  {Karpie}}, \bibinfo {author} {\bibfnamefont {K.}~\bibnamefont {Orginos}},
  \bibinfo {author} {\bibfnamefont {A.}~\bibnamefont {Radyushkin}}, \ and\
  \bibinfo {author} {\bibfnamefont {S.}~\bibnamefont {Zafeiropoulos}},\
  }\bibfield  {booktitle} {\emph {\bibinfo {booktitle} {{Proceedings, 35th
  International Symposium on Lattice Field Theory (Lattice 2017): Granada,
  Spain, June 18-24, 2017}}},\ }\href {\doibase 10.1051/epjconf/201817506032}
  {\bibfield  {journal} {\bibinfo  {journal} {EPJ Web Conf.}\ }\textbf
  {\bibinfo {volume} {175}},\ \bibinfo {pages} {06032} (\bibinfo {year}
  {2018}{\natexlab{a}})},\ \Eprint {http://arxiv.org/abs/1710.08288}
  {arXiv:1710.08288 [hep-lat]} \BibitemShut {NoStop}%
%%CITATION = ARXIV:1710.08288;%%
\bibitem [{\citenamefont {Karpie}\ \emph
  {et~al.}(2018{\natexlab{b}})\citenamefont {Karpie}, \citenamefont {Orginos},\
  and\ \citenamefont {Zafeiropoulos}}]{Karpie:2018zaz}%
  \BibitemOpen
  \bibfield  {author} {\bibinfo {author} {\bibfnamefont {J.}~\bibnamefont
  {Karpie}}, \bibinfo {author} {\bibfnamefont {K.}~\bibnamefont {Orginos}}, \
  and\ \bibinfo {author} {\bibfnamefont {S.}~\bibnamefont {Zafeiropoulos}},\
  }\href {\doibase 10.1007/JHEP11(2018)178} {\bibfield  {journal} {\bibinfo
  {journal} {JHEP}\ }\textbf {\bibinfo {volume} {11}},\ \bibinfo {pages} {178}
  (\bibinfo {year} {2018}{\natexlab{b}})},\ \Eprint
  {http://arxiv.org/abs/1807.10933} {arXiv:1807.10933 [hep-lat]} \BibitemShut
  {NoStop}%
%%CITATION = ARXIV:1807.10933;%%
\bibitem [{\citenamefont {Karpie}\ \emph {et~al.}(2019)\citenamefont {Karpie},
  \citenamefont {Orginos}, \citenamefont {Rothkopf},\ and\ \citenamefont
  {Zafeiropoulos}}]{Karpie:2019eiq}%
  \BibitemOpen
  \bibfield  {author} {\bibinfo {author} {\bibfnamefont {J.}~\bibnamefont
  {Karpie}}, \bibinfo {author} {\bibfnamefont {K.}~\bibnamefont {Orginos}},
  \bibinfo {author} {\bibfnamefont {A.}~\bibnamefont {Rothkopf}}, \ and\
  \bibinfo {author} {\bibfnamefont {S.}~\bibnamefont {Zafeiropoulos}},\ }\href
  {\doibase 10.1007/JHEP04(2019)057} {\bibfield  {journal} {\bibinfo  {journal}
  {JHEP}\ }\textbf {\bibinfo {volume} {04}},\ \bibinfo {pages} {057} (\bibinfo
  {year} {2019})},\ \Eprint {http://arxiv.org/abs/1901.05408} {arXiv:1901.05408
  [hep-lat]} \BibitemShut {NoStop}%
%%CITATION = ARXIV:1901.05408;%%
\bibitem [{\citenamefont {Jo\'o}\ \emph
  {et~al.}(2019{\natexlab{a}})\citenamefont {Jo\'o}, \citenamefont {Karpie},
  \citenamefont {Orginos}, \citenamefont {Radyushkin}, \citenamefont
  {Richards},\ and\ \citenamefont {Zafeiropoulos}}]{Joo:2019jct}%
  \BibitemOpen
  \bibfield  {author} {\bibinfo {author} {\bibfnamefont {B.}~\bibnamefont
  {Jo\'o}}, \bibinfo {author} {\bibfnamefont {J.}~\bibnamefont {Karpie}},
  \bibinfo {author} {\bibfnamefont {K.}~\bibnamefont {Orginos}}, \bibinfo
  {author} {\bibfnamefont {A.}~\bibnamefont {Radyushkin}}, \bibinfo {author}
  {\bibfnamefont {D.}~\bibnamefont {Richards}}, \ and\ \bibinfo {author}
  {\bibfnamefont {S.}~\bibnamefont {Zafeiropoulos}},\ }\href {\doibase
  10.1007/JHEP12(2019)081} {\bibfield  {journal} {\bibinfo  {journal} {JHEP}\
  }\textbf {\bibinfo {volume} {12}},\ \bibinfo {pages} {081} (\bibinfo {year}
  {2019}{\natexlab{a}})},\ \Eprint {http://arxiv.org/abs/1908.09771}
  {arXiv:1908.09771 [hep-lat]} \BibitemShut {NoStop}%
%%CITATION = ARXIV:1908.09771;%%
\bibitem [{\citenamefont {Jo\'o}\ \emph
  {et~al.}(2019{\natexlab{b}})\citenamefont {Jo\'o}, \citenamefont {Karpie},
  \citenamefont {Orginos}, \citenamefont {Radyushkin}, \citenamefont
  {Richards}, \citenamefont {Sufian},\ and\ \citenamefont
  {Zafeiropoulos}}]{Joo:2019bzr}%
  \BibitemOpen
  \bibfield  {author} {\bibinfo {author} {\bibfnamefont {B.}~\bibnamefont
  {Jo\'o}}, \bibinfo {author} {\bibfnamefont {J.}~\bibnamefont {Karpie}},
  \bibinfo {author} {\bibfnamefont {K.}~\bibnamefont {Orginos}}, \bibinfo
  {author} {\bibfnamefont {A.~V.}\ \bibnamefont {Radyushkin}}, \bibinfo
  {author} {\bibfnamefont {D.~G.}\ \bibnamefont {Richards}}, \bibinfo {author}
  {\bibfnamefont {R.~S.}\ \bibnamefont {Sufian}}, \ and\ \bibinfo {author}
  {\bibfnamefont {S.}~\bibnamefont {Zafeiropoulos}},\ }\href {\doibase
  10.1103/PhysRevD.100.114512} {\bibfield  {journal} {\bibinfo  {journal}
  {Phys. Rev.}\ }\textbf {\bibinfo {volume} {D100}},\ \bibinfo {pages} {114512}
  (\bibinfo {year} {2019}{\natexlab{b}})},\ \Eprint
  {http://arxiv.org/abs/1909.08517} {arXiv:1909.08517 [hep-lat]} \BibitemShut
  {NoStop}%
%%CITATION = ARXIV:1909.08517;%%
\bibitem [{\citenamefont {Radyushkin}(2019)}]{Radyushkin:2019mye}%
  \BibitemOpen
  \bibfield  {author} {\bibinfo {author} {\bibfnamefont {A.~V.}\ \bibnamefont
  {Radyushkin}},\ }\href@noop {} {\  (\bibinfo {year} {2019})},\ \Eprint
  {http://arxiv.org/abs/1912.04244} {arXiv:1912.04244 [hep-ph]} \BibitemShut
  {NoStop}%
%%CITATION = ARXIV:1912.04244;%%
\bibitem [{\citenamefont {Radyushkin}(2018)}]{Radyushkin:2018cvn}%
  \BibitemOpen
  \bibfield  {author} {\bibinfo {author} {\bibfnamefont {A.}~\bibnamefont
  {Radyushkin}},\ }\href {\doibase 10.1103/PhysRevD.98.014019} {\bibfield
  {journal} {\bibinfo  {journal} {Phys. Rev.}\ }\textbf {\bibinfo {volume}
  {D98}},\ \bibinfo {pages} {014019} (\bibinfo {year} {2018})},\ \Eprint
  {http://arxiv.org/abs/1801.02427} {arXiv:1801.02427 [hep-ph]} \BibitemShut
  {NoStop}%
%%CITATION = ARXIV:1801.02427;%%
\bibitem [{\citenamefont {Zhang}\ \emph {et~al.}(2018)\citenamefont {Zhang},
  \citenamefont {Chen},\ and\ \citenamefont {Monahan}}]{Zhang:2018ggy}%
  \BibitemOpen
  \bibfield  {author} {\bibinfo {author} {\bibfnamefont {J.-H.}\ \bibnamefont
  {Zhang}}, \bibinfo {author} {\bibfnamefont {J.-W.}\ \bibnamefont {Chen}}, \
  and\ \bibinfo {author} {\bibfnamefont {C.}~\bibnamefont {Monahan}},\ }\href
  {\doibase 10.1103/PhysRevD.97.074508} {\bibfield  {journal} {\bibinfo
  {journal} {Phys. Rev.}\ }\textbf {\bibinfo {volume} {D97}},\ \bibinfo {pages}
  {074508} (\bibinfo {year} {2018})},\ \Eprint
  {http://arxiv.org/abs/1801.03023} {arXiv:1801.03023 [hep-ph]} \BibitemShut
  {NoStop}%
%%CITATION = ARXIV:1801.03023;%%
\bibitem [{\citenamefont {Izubuchi}\ \emph {et~al.}(2018)\citenamefont
  {Izubuchi}, \citenamefont {Ji}, \citenamefont {Jin}, \citenamefont
  {Stewart},\ and\ \citenamefont {Zhao}}]{Izubuchi:2018srq}%
  \BibitemOpen
  \bibfield  {author} {\bibinfo {author} {\bibfnamefont {T.}~\bibnamefont
  {Izubuchi}}, \bibinfo {author} {\bibfnamefont {X.}~\bibnamefont {Ji}},
  \bibinfo {author} {\bibfnamefont {L.}~\bibnamefont {Jin}}, \bibinfo {author}
  {\bibfnamefont {I.~W.}\ \bibnamefont {Stewart}}, \ and\ \bibinfo {author}
  {\bibfnamefont {Y.}~\bibnamefont {Zhao}},\ }\href {\doibase
  10.1103/PhysRevD.98.056004} {\bibfield  {journal} {\bibinfo  {journal} {Phys.
  Rev.}\ }\textbf {\bibinfo {volume} {D98}},\ \bibinfo {pages} {056004}
  (\bibinfo {year} {2018})},\ \Eprint {http://arxiv.org/abs/1801.03917}
  {arXiv:1801.03917 [hep-ph]} \BibitemShut {NoStop}%
%%CITATION = ARXIV:1801.03917;%%
\bibitem [{\citenamefont {Altarelli}\ and\ \citenamefont
  {Parisi}(1977)}]{Altarelli:1977zs}%
  \BibitemOpen
  \bibfield  {author} {\bibinfo {author} {\bibfnamefont {G.}~\bibnamefont
  {Altarelli}}\ and\ \bibinfo {author} {\bibfnamefont {G.}~\bibnamefont
  {Parisi}},\ }\href {\doibase 10.1016/0550-3213(77)90384-4} {\bibfield
  {journal} {\bibinfo  {journal} {Nucl. Phys.}\ }\textbf {\bibinfo {volume}
  {B126}},\ \bibinfo {pages} {298} (\bibinfo {year} {1977})}\BibitemShut
  {NoStop}%
%%CITATION = NUPHA,B126,298;%%
\bibitem [{\citenamefont {Yoon}\ \emph {et~al.}(2017)\citenamefont {Yoon} \emph
  {et~al.}}]{Yoon:2016jzj}%
  \BibitemOpen
  \bibfield  {author} {\bibinfo {author} {\bibfnamefont {B.}~\bibnamefont
  {Yoon}} \emph {et~al.},\ }\href {\doibase 10.1103/PhysRevD.95.074508}
  {\bibfield  {journal} {\bibinfo  {journal} {Phys. Rev.}\ }\textbf {\bibinfo
  {volume} {D95}},\ \bibinfo {pages} {074508} (\bibinfo {year} {2017})},\
  \Eprint {http://arxiv.org/abs/1611.07452} {arXiv:1611.07452 [hep-lat]}
  \BibitemShut {NoStop}%
%%CITATION = ARXIV:1611.07452;%%
\bibitem [{\citenamefont {Allton}\ \emph {et~al.}(1993)\citenamefont {Allton}
  \emph {et~al.}}]{Allton:1993wc}%
  \BibitemOpen
  \bibfield  {author} {\bibinfo {author} {\bibfnamefont {C.~R.}\ \bibnamefont
  {Allton}} \emph {et~al.} (\bibinfo {collaboration} {UKQCD}),\ }\href
  {\doibase 10.1103/PhysRevD.47.5128} {\bibfield  {journal} {\bibinfo
  {journal} {Phys. Rev.}\ }\textbf {\bibinfo {volume} {D47}},\ \bibinfo {pages}
  {5128} (\bibinfo {year} {1993})},\ \Eprint
  {http://arxiv.org/abs/hep-lat/9303009} {arXiv:hep-lat/9303009 [hep-lat]}
  \BibitemShut {NoStop}%
%%CITATION = HEP-LAT/9303009;%%
\bibitem [{\citenamefont {Bali}\ \emph {et~al.}(2016)\citenamefont {Bali},
  \citenamefont {Lang}, \citenamefont {Musch},\ and\ \citenamefont
  {Sch{\"a}fer}}]{Bali:2016lva}%
  \BibitemOpen
  \bibfield  {author} {\bibinfo {author} {\bibfnamefont {G.~S.}\ \bibnamefont
  {Bali}}, \bibinfo {author} {\bibfnamefont {B.}~\bibnamefont {Lang}}, \bibinfo
  {author} {\bibfnamefont {B.~U.}\ \bibnamefont {Musch}}, \ and\ \bibinfo
  {author} {\bibfnamefont {A.}~\bibnamefont {Sch{\"a}fer}},\ }\href {\doibase
  10.1103/PhysRevD.93.094515} {\bibfield  {journal} {\bibinfo  {journal} {Phys.
  Rev.}\ }\textbf {\bibinfo {volume} {D93}},\ \bibinfo {pages} {094515}
  (\bibinfo {year} {2016})},\ \Eprint {http://arxiv.org/abs/1602.05525}
  {arXiv:1602.05525 [hep-lat]} \BibitemShut {NoStop}%
%%CITATION = ARXIV:1602.05525;%%
\bibitem [{\citenamefont {Bouchard}\ \emph {et~al.}(2017)\citenamefont
  {Bouchard}, \citenamefont {Chang}, \citenamefont {Kurth}, \citenamefont
  {Orginos},\ and\ \citenamefont {Walker-Loud}}]{Bouchard:2016heu}%
  \BibitemOpen
  \bibfield  {author} {\bibinfo {author} {\bibfnamefont {C.}~\bibnamefont
  {Bouchard}}, \bibinfo {author} {\bibfnamefont {C.~C.}\ \bibnamefont {Chang}},
  \bibinfo {author} {\bibfnamefont {T.}~\bibnamefont {Kurth}}, \bibinfo
  {author} {\bibfnamefont {K.}~\bibnamefont {Orginos}}, \ and\ \bibinfo
  {author} {\bibfnamefont {A.}~\bibnamefont {Walker-Loud}},\ }\href {\doibase
  10.1103/PhysRevD.96.014504} {\bibfield  {journal} {\bibinfo  {journal} {Phys.
  Rev.}\ }\textbf {\bibinfo {volume} {D96}},\ \bibinfo {pages} {014504}
  (\bibinfo {year} {2017})},\ \Eprint {http://arxiv.org/abs/1612.06963}
  {arXiv:1612.06963 [hep-lat]} \BibitemShut {NoStop}%
%%CITATION = ARXIV:1612.06963;%%
\bibitem [{\citenamefont {Accardi}\ \emph {et~al.}(2016)\citenamefont
  {Accardi}, \citenamefont {Brady}, \citenamefont {Melnitchouk}, \citenamefont
  {Owens},\ and\ \citenamefont {Sato}}]{CJ}%
  \BibitemOpen
  \bibfield  {author} {\bibinfo {author} {\bibfnamefont {A.}~\bibnamefont
  {Accardi}}, \bibinfo {author} {\bibfnamefont {L.~T.}\ \bibnamefont {Brady}},
  \bibinfo {author} {\bibfnamefont {W.}~\bibnamefont {Melnitchouk}}, \bibinfo
  {author} {\bibfnamefont {J.~F.}\ \bibnamefont {Owens}}, \ and\ \bibinfo
  {author} {\bibfnamefont {N.}~\bibnamefont {Sato}},\ }\href {\doibase
  10.1103/PhysRevD.93.114017} {\bibfield  {journal} {\bibinfo  {journal} {Phys.
  Rev.}\ }\textbf {\bibinfo {volume} {D93}},\ \bibinfo {pages} {114017}
  (\bibinfo {year} {2016})},\ \Eprint {http://arxiv.org/abs/1602.03154}
  {arXiv:1602.03154 [hep-ph]} \BibitemShut {NoStop}%
%%CITATION = ARXIV:1602.03154;%%
\bibitem [{\citenamefont {Martin}\ \emph {et~al.}(2009)\citenamefont {Martin},
  \citenamefont {Stirling}, \citenamefont {Thorne},\ and\ \citenamefont
  {Watt}}]{Martin:2009iq}%
  \BibitemOpen
  \bibfield  {author} {\bibinfo {author} {\bibfnamefont {A.~D.}\ \bibnamefont
  {Martin}}, \bibinfo {author} {\bibfnamefont {W.~J.}\ \bibnamefont
  {Stirling}}, \bibinfo {author} {\bibfnamefont {R.~S.}\ \bibnamefont
  {Thorne}}, \ and\ \bibinfo {author} {\bibfnamefont {G.}~\bibnamefont
  {Watt}},\ }\href {\doibase 10.1140/epjc/s10052-009-1072-5} {\bibfield
  {journal} {\bibinfo  {journal} {Eur. Phys. J.}\ }\textbf {\bibinfo {volume}
  {C63}},\ \bibinfo {pages} {189} (\bibinfo {year} {2009})},\ \Eprint
  {http://arxiv.org/abs/0901.0002} {arXiv:0901.0002 [hep-ph]} \BibitemShut
  {NoStop}%
%%CITATION = ARXIV:0901.0002;%%
\bibitem [{\citenamefont {Ball}\ \emph {et~al.}(2017)\citenamefont {Ball} \emph
  {et~al.}}]{Ball:2017nwa}%
  \BibitemOpen
  \bibfield  {author} {\bibinfo {author} {\bibfnamefont {R.~D.}\ \bibnamefont
  {Ball}} \emph {et~al.} (\bibinfo {collaboration} {NNPDF}),\ }\href {\doibase
  10.1140/epjc/s10052-017-5199-5} {\bibfield  {journal} {\bibinfo  {journal}
  {Eur. Phys. J.}\ }\textbf {\bibinfo {volume} {C77}},\ \bibinfo {pages} {663}
  (\bibinfo {year} {2017})},\ \Eprint {http://arxiv.org/abs/1706.00428}
  {arXiv:1706.00428 [hep-ph]} \BibitemShut {NoStop}%
%%CITATION = ARXIV:1706.00428;%%
\bibitem [{\citenamefont {Cichy}\ \emph {et~al.}(2019)\citenamefont {Cichy},
  \citenamefont {Del~Debbio},\ and\ \citenamefont {Giani}}]{Cichy:2019ebf}%
  \BibitemOpen
  \bibfield  {author} {\bibinfo {author} {\bibfnamefont {K.}~\bibnamefont
  {Cichy}}, \bibinfo {author} {\bibfnamefont {L.}~\bibnamefont {Del~Debbio}}, \
  and\ \bibinfo {author} {\bibfnamefont {T.}~\bibnamefont {Giani}},\ }\href
  {\doibase 10.1007/JHEP10(2019)137} {\bibfield  {journal} {\bibinfo  {journal}
  {JHEP}\ }\textbf {\bibinfo {volume} {10}},\ \bibinfo {pages} {137} (\bibinfo
  {year} {2019})},\ \Eprint {http://arxiv.org/abs/1907.06037} {arXiv:1907.06037
  [hep-ph]} \BibitemShut {NoStop}%
%%CITATION = ARXIV:1907.06037;%%
\bibitem [{\citenamefont {Constantinou}(2015)}]{Constantinou:2014tga}%
  \BibitemOpen
  \bibfield  {author} {\bibinfo {author} {\bibfnamefont {M.}~\bibnamefont
  {Constantinou}},\ }\bibfield  {booktitle} {\emph {\bibinfo {booktitle}
  {{Proceedings, 32nd International Symposium on Lattice Field Theory (Lattice
  2014): Brookhaven, NY, USA, June 23-28, 2014}}},\ }\href {\doibase
  10.22323/1.214.0001} {\bibfield  {journal} {\bibinfo  {journal} {PoS}\
  }\textbf {\bibinfo {volume} {LATTICE2014}},\ \bibinfo {pages} {001} (\bibinfo
  {year} {2015})},\ \Eprint {http://arxiv.org/abs/1411.0078} {arXiv:1411.0078
  [hep-lat]} \BibitemShut {NoStop}%
%%CITATION = ARXIV:1411.0078;%%
\bibitem [{\citenamefont {{J\"{u}lich Supercomputing Centre}}(2018)}]{jureca}%
  \BibitemOpen
  \bibfield  {author} {\bibinfo {author} {\bibnamefont {{J\"{u}lich
  Supercomputing Centre}}},\ }\href {\doibase 10.17815/jlsrf-4-121-1}
  {\bibfield  {journal} {\bibinfo  {journal} {Journal of large-scale research
  facilities}\ }\textbf {\bibinfo {volume} {4}} (\bibinfo {year} {2018}),\
  10.17815/jlsrf-4-121-1}\BibitemShut {NoStop}%
\bibitem [{\citenamefont {Towns}\ \emph {et~al.}(2014)\citenamefont {Towns},
  \citenamefont {Cockerill}, \citenamefont {Dahan}, \citenamefont {Foster},
  \citenamefont {Gaither}, \citenamefont {Grimshaw}, \citenamefont {Hazlewood},
  \citenamefont {Lathrop}, \citenamefont {Lifka}, \citenamefont {Peterson},
  \citenamefont {Roskies}, \citenamefont {Scott},\ and\ \citenamefont
  {Wilkins-Diehr}}]{xsede}%
  \BibitemOpen
  \bibfield  {author} {\bibinfo {author} {\bibfnamefont {J.}~\bibnamefont
  {Towns}}, \bibinfo {author} {\bibfnamefont {T.}~\bibnamefont {Cockerill}},
  \bibinfo {author} {\bibfnamefont {M.}~\bibnamefont {Dahan}}, \bibinfo
  {author} {\bibfnamefont {I.}~\bibnamefont {Foster}}, \bibinfo {author}
  {\bibfnamefont {K.}~\bibnamefont {Gaither}}, \bibinfo {author} {\bibfnamefont
  {A.}~\bibnamefont {Grimshaw}}, \bibinfo {author} {\bibfnamefont
  {V.}~\bibnamefont {Hazlewood}}, \bibinfo {author} {\bibfnamefont
  {S.}~\bibnamefont {Lathrop}}, \bibinfo {author} {\bibfnamefont
  {D.}~\bibnamefont {Lifka}}, \bibinfo {author} {\bibfnamefont {G.~D.}\
  \bibnamefont {Peterson}}, \bibinfo {author} {\bibfnamefont {R.}~\bibnamefont
  {Roskies}}, \bibinfo {author} {\bibfnamefont {J.}~\bibnamefont {Scott}}, \
  and\ \bibinfo {author} {\bibfnamefont {N.}~\bibnamefont {Wilkins-Diehr}},\
  }\href {\doibase 10.1109/MCSE.2014.80} {\bibfield  {journal} {\bibinfo
  {journal} {Computing in Science \& Engineering}\ }\textbf {\bibinfo {volume}
  {16}},\ \bibinfo {pages} {62} (\bibinfo {year} {2014})}\BibitemShut {NoStop}%
\bibitem [{\citenamefont {Edwards}\ and\ \citenamefont
  {Joo}(2005)}]{Edwards:2004sx}%
  \BibitemOpen
  \bibfield  {author} {\bibinfo {author} {\bibfnamefont {R.~G.}\ \bibnamefont
  {Edwards}}\ and\ \bibinfo {author} {\bibfnamefont {B.}~\bibnamefont {Joo}}
  (\bibinfo {collaboration} {SciDAC, LHPC, UKQCD}),\ }\href {\doibase
  10.1016/j.nuclphysbps.2004.11.254} {\bibfield  {journal} {\bibinfo  {journal}
  {Nucl. Phys. B Proc. Suppl.}\ }\textbf {\bibinfo {volume} {140}},\ \bibinfo
  {pages} {832} (\bibinfo {year} {2005})},\ \Eprint
  {http://arxiv.org/abs/hep-lat/0409003} {arXiv:hep-lat/0409003} \BibitemShut
  {NoStop}%
\bibitem [{\citenamefont {Clark}\ \emph {et~al.}(2010)\citenamefont {Clark},
  \citenamefont {Babich}, \citenamefont {Barros}, \citenamefont {Brower},\ and\
  \citenamefont {Rebbi}}]{Clark:2009wm}%
  \BibitemOpen
  \bibfield  {author} {\bibinfo {author} {\bibfnamefont {M.}~\bibnamefont
  {Clark}}, \bibinfo {author} {\bibfnamefont {R.}~\bibnamefont {Babich}},
  \bibinfo {author} {\bibfnamefont {K.}~\bibnamefont {Barros}}, \bibinfo
  {author} {\bibfnamefont {R.}~\bibnamefont {Brower}}, \ and\ \bibinfo {author}
  {\bibfnamefont {C.}~\bibnamefont {Rebbi}},\ }\href {\doibase
  10.1016/j.cpc.2010.05.002} {\bibfield  {journal} {\bibinfo  {journal}
  {Comput. Phys. Commun.}\ }\textbf {\bibinfo {volume} {181}},\ \bibinfo
  {pages} {1517} (\bibinfo {year} {2010})},\ \Eprint
  {http://arxiv.org/abs/0911.3191} {arXiv:0911.3191 [hep-lat]} \BibitemShut
  {NoStop}%
\bibitem [{\citenamefont {Babich}\ \emph {et~al.}(2010)\citenamefont {Babich},
  \citenamefont {Clark},\ and\ \citenamefont {Joo}}]{Babich:2010mu}%
  \BibitemOpen
  \bibfield  {author} {\bibinfo {author} {\bibfnamefont {R.}~\bibnamefont
  {Babich}}, \bibinfo {author} {\bibfnamefont {M.~A.}\ \bibnamefont {Clark}}, \
  and\ \bibinfo {author} {\bibfnamefont {B.}~\bibnamefont {Joo}},\ }in\
  \href@noop {} {\emph {\bibinfo {booktitle} {{SC 10 (Supercomputing 2010)}}}}\
  (\bibinfo {year} {2010})\ \Eprint {http://arxiv.org/abs/1011.0024}
  {arXiv:1011.0024 [hep-lat]} \BibitemShut {NoStop}%
\bibitem [{\citenamefont {Jo{\'o}}\ \emph {et~al.}(2016)\citenamefont
  {Jo{\'o}}, \citenamefont {Kalamkar}, \citenamefont {Kurth}, \citenamefont
  {Vaidyanathan},\ and\ \citenamefont {Walden}}]{QPhiX2}%
  \BibitemOpen
  \bibfield  {author} {\bibinfo {author} {\bibfnamefont {B.}~\bibnamefont
  {Jo{\'o}}}, \bibinfo {author} {\bibfnamefont {D.~D.}\ \bibnamefont
  {Kalamkar}}, \bibinfo {author} {\bibfnamefont {T.}~\bibnamefont {Kurth}},
  \bibinfo {author} {\bibfnamefont {K.}~\bibnamefont {Vaidyanathan}}, \ and\
  \bibinfo {author} {\bibfnamefont {A.}~\bibnamefont {Walden}},\ }in\ \href
  {\doibase 10.1007/978-3-319-46079-6_30} {\emph {\bibinfo {booktitle} {{High
  Performance Computing: ISC High Performance 2016 International Workshops,
  ExaComm, E-MuCoCoS, HPC-IODC, IXPUG, IWOPH, $P^3$MA, VHPC, WOPSSS, Frankfurt,
  Germany, June 19--23, 2016, Revised Selected Papers}}}},\ \bibinfo {editor}
  {edited by\ \bibinfo {editor} {\bibfnamefont {M.}~\bibnamefont {Taufer}},
  \bibinfo {editor} {\bibfnamefont {B.}~\bibnamefont {Mohr}}, \ and\ \bibinfo
  {editor} {\bibfnamefont {J.~M.}\ \bibnamefont {Kunkel}}}\ (\bibinfo
  {publisher} {Springer International Publishing},\ \bibinfo {address} {Cham},\
  \bibinfo {year} {2016})\ pp.\ \bibinfo {pages} {415--427}\BibitemShut
  {NoStop}%
\bibitem [{\citenamefont {Bhat}\ \emph {et~al.}(2020)\citenamefont {Bhat},
  \citenamefont {Cichy}, \citenamefont {Constantinou},\ and\ \citenamefont
  {Scapellato}}]{Bhat:2020ktg}%
  \BibitemOpen
  \bibfield  {author} {\bibinfo {author} {\bibfnamefont {M.}~\bibnamefont
  {Bhat}}, \bibinfo {author} {\bibfnamefont {K.}~\bibnamefont {Cichy}},
  \bibinfo {author} {\bibfnamefont {M.}~\bibnamefont {Constantinou}}, \ and\
  \bibinfo {author} {\bibfnamefont {A.}~\bibnamefont {Scapellato}},\
  }\href@noop {} {\  (\bibinfo {year} {2020})},\ \Eprint
  {http://arxiv.org/abs/2005.02102} {arXiv:2005.02102 [hep-lat]} \BibitemShut
  {NoStop}%
\end{thebibliography}%

\end{document}